\documentclass[preprint2,letterpaper]{aastex}
\usepackage{amsmath,url}

\newcommand\HII{\ion{H}{2}}
\newcommand\etal{{\it et al.}}
\newcommand\eg{{\it e.g.}}
\newcommand\ie{{\it i.e.}}
\newcommand\lfir{\ensuremath{L_{\rm FIR}}}

\newcommand\baruph{\ensuremath{{\bar U}_{\rm ph}}}
\newcommand\beq{\ensuremath{B_{\rm eq}}}
\newcommand\bmin{\ensuremath{B_{\rm min}}}
\newcommand\tsys{\ensuremath{T_{\rm sys}}}
\newcommand\uv{$u$-$v$}

\newcommand\tnu{\ensuremath{\nu_g}}
\newcommand\srcs{\ensuremath{\sqrt{\chi_r^2}}}
\newcommand\tff{\ensuremath{\tau_\mathrm{ff}}}
\newcommand\heff{\ensuremath{h_\mathrm{eff}}}
\newlength{\otofs}
\newcommand\otd{\setlength{\parindent}{-1\otofs}}
\defcitealias{t06}{T+06}
\newcommand\thomp{\citetalias{t06}}

\slugcomment{\today}
\shorttitle{Broadband Spectra with ATA}
\shortauthors{Williams \& Bower}

\begin{document}

\title{Evaluating the Calorimeter Model with \\
Broadband, Continuous Spectra of Starburst Galaxies \\
Observed with the Allen Telescope Array}
\author{Peter~K.~G.~Williams}
\affil{University of California, Berkeley, Department of
  Astronomy}
\affil{601 Campbell Hall \# 3411, Berkeley, CA 94720-3411}
\email{pwilliams@astro.berkeley.edu}
\and
\author{Geoffrey~C.~Bower}
\affil{University of California, Berkeley, Department of
  Astronomy}
\affil{601 Campbell Hall \# 3411, Berkeley, CA 94720-3411}
\email{gbower@astro.berkeley.edu}

\begin{abstract}
  Although the relationship between the far-infrared and cm-wave radio
  luminosities of normal galaxies is one of the most striking
  correlations in astronomy, a solid understanding of its physical
  basis is lacking. In one interpretation, the ``calorimeter model,''
  rapid synchrotron cooling of cosmic ray electrons is essential in
  reproducing the observed linear relationship. Observed radio
  spectra, however, are shallower than what is expected of cooled
  synchrotron emission. In Thompson \etal\ (2006), a simple
  parameterized model is presented to explain how relatively shallow
  observed spectra might arise even in the presence of rapid
  synchrotron cooling by accounting for ionization losses and other
  cooling mechanisms. During the commissioning of the 42-element Allen
  Telescope Array, we observed the starburst galaxies M82, NGC~253,
  and Arp~220 at frequencies ranging from 1 to 7 GHz, obtaining
  unprecedented broadband continuous radio spectra of these sources.
  We combine our observations with high-frequency data from the
  literature to separate the spectra into thermal and nonthermal
  components. The nonthermal components all steepen in the cm-wave
  regime and cannot be well-modeled as simple power laws.  The model
  of Thompson \etal\ is consistent with our M82 results when plausible
  parameters are chosen, and our results in fact significantly shrink
  the space of allowed model parameters. The model is only marginally
  consistent with our NGC~253 data. Assuming the Thompson
  \etal\ model, a steep electron energy injection index of $p = -2.5$
  is ruled out in M82 and NGC~253 to $>$99\% confidence. We describe in
  detail the observing procedures, calibration methods, analysis, and
  consistency checks used for broadband spectral observations with the
  Allen Telescope Array.
\end{abstract}

\section{Introduction}
\label{s:intro}

The correlation between the far-infrared (FIR) and centimeter-wave
radio emission of normal galaxies has been recognized for almost four
decades \citep{vdk71}. It holds over more than four orders of
magnitude in \lfir\ \citep{y01}, is constant out to at least $z = 1$
\citep{a+04}, and is observed in a large variety of galaxies
\citep{c92}. The near-linearity \citep{y01} and universality of the
FIR-radio correlation strongly suggest that it is intimately linked to
the fundamental astrophysics of galaxies.

The physical underpinning of the correlation is broadly understood to
be the formation of massive stars, which can believably be associated
with the dominant contributions to both the FIR and continuum cm-wave
radio emission of galaxies \citep{hp75,wk88}. The FIR contribution is
due to the hot dust envelopes surrounding these stars, which are
energized by the copious amounts of Lyman continuum radiation that
massive stars emit. The radio contribution includes two components.
The first is synchrotron emission from relativistic cosmic ray (CR)
electrons spiraling in the interstellar magnetic field. Such electrons
can be accelerated in the supernova (SN) shocks that occur when
massive stars die. The second, less significant, component is
free-free emission, which predominantly comes from \HII\ regions.
These too are associated with massive stars.

This simple picture, however, neglects many important details. The
timescales and dynamics of the radio and FIR emission are very
different: CR electrons radiate over a timescale of $10^4$-$10^7$ yr
(see below), during which they diffuse through the galaxy, are
possibly advected out of it via a galactic wind \citep{bmv91}, and
cool via a variety of mechanisms, including inverse Compton (IC)
scattering, ionization, free-free emission, and adiabatic expansion.
Meanwhile, the FIR emission is prompt and comprised of comparable warm
($\sim$40 K) and cool (``cirrus,'' $\sim$20 K) dust components, the
latter of which is primarily heated by the interstellar radiation
field, not newly-formed stars \citep{ws87,bp88,fac88}.

Much theoretical and observational effort has been expended in the
cause of understanding the FIR-radio correlation in detail, (\eg\ 
\citet{dkww85,hsr85,ceaf88,hdwhp88,de89,pd92,lvx96}). An influential
model has been that of \citet{v89}, which presents a theory for
understanding the FIR-radio correlation in the case that CR electrons
are accelerated only by SNe and cool completely within a galaxy. In
this situation, the galaxy acts as a ``calorimeter'' with regard to
the energy injected into CR electrons. For the FIR-radio correlation
to hold, the synchrotron luminosity per unit energy injected into CR
electrons must be a universal constant. Taking the primary cooling
mechanisms to be IC scattering and synchrotron, it is found that $\bar
U_B / (\baruph + \bar U_B)$ must be nearly a constant value, where
$\bar U_B = \langle B^2/8\pi\rangle$ is the mean energy density of the
interstellar magnetic field and $\baruph$ is the mean energy density
of the interstellar radiation field.

Because the synchrotron cooling time scales as $B^{-3/2}$
\citep[hereafter T06]{t06}, the applicability of the calorimeter
model depends on the strength of the interstellar magnetic field. In a
stronger field, the calorimeter model is more likely to hold as CR
electrons will cool more rapidly and have less chance to escape the
galactic disk before radiating all of their endowed energy. This is
precisely the scenario suggested in \thomp, where it is argued that
starforming galaxies have much stronger magnetic fields than commonly
assumed. The work assesses the applicability of the calorimeter model
by considering the timescales for synchrotron cooling and escape from
the galactic disk by CR electrons. Using the usual assumptions, the
two are comparable; using the stronger magnetic field that \thomp\ argues
for, the synchrotron cooling timescale is much shorter, favoring the
assumptions of the calorimeter model.

\thomp\ specifically argues that in starburst galaxies the
magnetic field is dynamically relevant: its energy density is
comparable to the hydrostatic pressure of the ISM, or
$$
B \approx \beq = (8\pi^2G)^{1/2}\Sigma_g \approx 
2.3 \left(\frac{\Sigma_g}{1\textrm{ g cm}^{-2}}\right)\textrm{ mG},
$$
where $\Sigma_g$ is the surface density of the galactic gas disk.
This is in contrast to the classic ``minimum-energy'' estimate
\citep{b56,l94},
$$
B \approx \bmin \approx 9.5\, \eta^{2/7}
\left(\frac{L_\nu}{10^{23}\textrm{ W Hz}^{-1}}\right)^{2/7}
\left(\frac{V}{1\textrm{ pc}^3}\right)^{-2/7}
\left(\frac{\nu}{1\textrm{ GHz}}\right)^{1/7}\textrm{ mG},
$$
where $\eta$ is the ratio of total CR energy to energy in the CR
electrons alone, $L_\nu$ is the synchrotron luminosity, $V$ is the
estimated volume of the emitting region, and $\nu$ is the observing
frequency. For the parameter values commonly associated with starburst
galaxies, $\beq \gg \bmin$ \citepalias[fig.\ 1]{t06}.

An important observational constraint on the argument of \thomp\ is
the shape of the galactic nonthermal spectrum. It is typically argued
that galactic spectra are insufficiently steep to be compatible with
the calorimeter model: observations indicate $\alpha \approx -0.8$
($S_\nu \propto \nu^\alpha$; \citet{c92,nkw97}) rather than $\alpha
\approx -1$ as would be expected from the combination of SN shock
physics \citep{be87} and spectral steepening caused by strong cooling
\citep{c92}. \thomp\ contends that more detailed analysis mitigates
this issue. As CR electron energy decreases, the timescales of
ionization losses and free-free emission are, respectively, shorter
and unchanged, so both of these processes will tend to flatten
observed spectra. In \thomp\ simple steady-state simulations were
performed to quantify this effect (cf.\ its fig.\ 3 and \S4.3).  These
simulations lead to the general prediction that the nonthermal radio
spectra of starburst galaxies should steepen with increasing
frequency, asymptoting to $\alpha \lesssim -1$.

Most spectral index measurements are based on only a handful of
fluxes. To accurately measure the steepening of a nonthermal spectrum,
many more data are needed, because both the steepening and the thermal
contribution must be determined. The work of \citet{dbg88} and
\citet{nkw97} shows that determination of the thermal fraction is
difficult and requires high-quality high-frequency data, even if a
simple power-law nonthermal spectrum is assumed. As \citet{c92}
points out, the thermal fraction and nonthermal steepening can also be
highly degenerate (cf.\ his Figure~6). One must sample a large number
of points over a wide frequency range in uniform conditions to
effectively constrain these parameters.

Though it is still being commissioned, the Allen Telescope Array (ATA;
\citet{w+09}), with its continuous frequency coverage of 0.5 GHz to 10
GHz, is well-suited to testing the predictions of \thomp\ by observing
high-resolution, continuous spectra of starburst galaxies.  Section
\ref{s:obs} describes observations of M82, NGC~253, and Arp~220 made
with the ATA in the fall and winter of 2008. In \S\ref{s:redux} our
data reduction methods are described along with results for standard
calibrator sources. Section \ref{s:model} describes our modeling of
the observed spectra to extract thermal and nonthermal contributions.
Finally, in \S\ref{s:disc} we discuss our results and in \S\ref{s:conc}
we present our conclusions.

\section{Observations}
\label{s:obs}

We observed the starforming galaxies M82, NGC~253, and Arp~220 with
the Allen Telescope Array on 
seven nights over the course of 2008 September to 2009 June.
The ATA is a large-$N$, small-$D$ (LNSD)
cm-wave array located in Hat Creek, California, and is a joint project
of the SETI\footnote{Search for Extraterrestrial Intelligence}
Institute of Mountain View, CA and the Radio Astronomy Laboratory of
the University of California, Berkeley. The scientific motivators and
technical capabilities of the ATA are described in \citet{w+09}. At
the time of the observations the ATA was undergoing commissioning,
with many aspects of the array being actively developed.  Forty-two
antennas and one correlator had been installed at the site.
Two-thirds of the antennas were typically available for use during our
observations.

We used the continuous frequency coverage of the ATA to observe the
galaxies at almost a decade of sky frequency in a sequence of snapshot
observations. 
The observing frequencies ranged between 1.0 GHz and 7.0 GHz with a
spacing of 100 MHz, skipping some frequencies due to known
radiofrequency interference (RFI) or low priority. The primary beam
sizes of the ATA at 1.0 GHz and 7.0 GHz are $\sim$210\,\arcmin\ and
$\sim$30\,\arcmin, respectively, and the synthesized beam sizes are
typically $\sim$250\arcsec\ and $\sim$35\arcsec. The snapshots lasted
between 60 and 130 seconds, with the correlator dump time always being
10 seconds. All observations used a bandwidth of 104.9 MHz and 1024
correlator channels and sampled both vertical and horizontal linear
polarizations.  Observations of the science targets were interleaved
with observations of the calibrators 3C~48, 3C~147, and 3C~286. These
calibrators are compact to the ATA and have up-to-date broadband flux
density models available from the
VLA.
\footnote{\url{http://purl.oclc.org/net/pkgwpub/vla-analytic-cal-models}}
Because of the need to set the flux density scale at numerous
frequencies and the desire to obtain a large amount of data for
sanity-checking, a large portion of the observing time was dedicated
to calibrator observations. The total integration time on the
calibrators was ten hours, compared to nine hours spent on the target
galaxies. An observing strategy that gathered fewer redundant data
could spend approximately half as much time observing calibrators as
observing science targets.

\subsection{Details of observing methods}

Because the observations presented herein use a novel technique on a
novel instrument, we shall describe our methods in detail. Each
night was divided into several segments during which a subset of the
targets and calibrators would be observed as a group. The typical
duration of a segment was a few hours. Over the course of each segment, an
observing script would repeatedly slew to each of the selected sources
in turn, observing each at several frequencies before moving on to the
next, until the segment's allotted time had elapsed.  This scheme
attempted to balance the desire to obtain calibrator observations
temporally close to the science observations with the desire to not
spend too much time moving from one target to the next.  Most segments
included observations at 1.4 GHz and 5.0 GHz to enable checks of
repeatability and stability. The observations presented in this paper
are summarized in Table~\ref{t:obs}.

Observing over a large range of frequencies requires not just a large
number of calibrator scans but also reconfiguration of the ATA
hardware. In particular, the sensitivity of the log-periodic ATA feeds
is a function of position along the axis of their pyramidal structure,
with the point of peak sensitivity approaching the apex of the pyramid
with increasing frequency. To allow for maximal performance, the ATA
feeds can be focused by moving them in a piston motion along the
normal to each antenna's focal plane \citep[see][fig.~4]{w+09}. The
optimal focus positions have been determined theoretically and
calibrated so that an ATA observer can specify the desired focus
position in frequency space. The focusing operation, however, is slow.
In our work, before each observation we set the feed focus frequency
to the observing frequency rounded up to an integral number of GHz
(\eg, 4.0, 4.4, and 4.9 GHz all map to 5 GHz).  This scheme was chosen
because the sensitivity of the feed to frequencies above the nominal
focus frequency drops off more rapidly than the sensitivity to
frequencies below it.  Within each segment we chose observing
frequencies that lay within a single GHz span to avoid superfluous
refocus operations. The focusing and other time-consuming operations
are parallelized as much as possible to conserve observing time.

Because neither the feed sensitivity nor the noise power entering the
ATA are spectrally flat, when the focus position of a feed changes,
the total power entering the analog system also changes. Just before
digitization, the downmixed and filtered analog signal is passed
through an ``attemplifier'' (attenuator/amplifier) stage to adjust the
incoming power for optimal quantization \citep{b07}. The amount of
attenuation or amplification must be explicitly configured, but the best
settings can be calibrated automatically during observations. Clearly,
any two observations that are to be compared or used for flux
calibration must be made with the same attemplifier
settings. Furthermore, the attemplifier calibration step is also
somewhat slow ($\sim$60 s). In these observations, snapshots made at
similar frequencies shared attemplifier settings. Specifically, one
set of attemplifier settings was used for each integral GHz span (\eg,
$4 \le \nu_\mathrm{GHz} < 5$). The first time any frequency in a span
is observed, the attemplifier settings are calibrated for that
frequency and saved; for subsequent observations during the same
segment, the saved settings are reapplied.

As an example of all of these considerations, consider segment \#12,
as detailed in Table~\ref{t:oscans}. This segment was scheduled for
observation from 03:42 UT to 05:12 on 2008 Oct 15. (All dates and
times are in UT.) The selected sources were 3C~48, 3C~147, and M82.
The timing and (short) duration of this segment were set by the
visibility of the science targets to the ATA: Arp~220 set just before
this segment began, and NGC~253 rose just after it ended. Because M82
is circumpolar at the ATA's latitude, the other two targets were
prioritized whenever they were visible. The selected observing
frequencies were 1.4, 2.4, 2.5, 4.3, 4.7, and 5.0 GHz. As described
above, the observing script slewed to the sources in turn and observed
each at several frequencies. The frequencies were chosen and grouped
so as to reduce the amount of time spent adjusting the feed focus
positions and attemplifier settings. At 04:08, the script would have
observed 3C~147 at the trio of lower frequencies if possible, but that
source had not yet risen. Instead, the script skipped those
observations and looped back to observe 3C~48 at the trio of higher
frequencies.

Including time lost due to slewing, hardware initialization, software
overhead, and so on, we made
$\sim$18 snapshots per hour while observing, integrating for 38\% of
the 53 hours spent using telescope resources during the observations
presented in this paper. The efficiency of future observations will be
higher: we have made an unusually large number of calibrator
observations and have made no effort to reduce slew time between
sources. Furthermore, our observations were made before the
installation of a second independently-tunable correlator at Hat
Creek, which doubles the observational throughput.

\section{Data Reduction}
\label{s:redux}

The data were reduced with a combination of the analysis package
MIRIAD
\footnote{\url{http://purl.oclc.org/net/pkgwpub/miriad}} \citep{stw95}
and custom Python
\footnote{\url{http://www.python.org/}} scripts based on the
NumPy and IPython packages. Over the course of this project and other
work on the ATA, significant improvements have been made to MIRIAD to
ease its compilation, improve its ability to handle large datasets,
and fix various bugs. A special module was written to allow easy
management of and access to MIRIAD datasets in the Python language.%
\footnote{\url{http://purl.oclc.org/net/pkgwpub/miriad-python}}

The two orthogonally-polarized signals that are received by the ATA
feed elements are amplified by separate low-noise amplifiers (LNAs)
and subsequently treated as entirely separate signals. The ATA
correlator at the time of the observations accepted 64 such
antenna-polarization signals, or ``antpols'' for short, making it
capable of correlating 32 antennas with full-polarization coverage, 64
antennas with single-polarization coverage, or, as encountered in
practice, an intermediate case. Each ATA antenna is identified by a
small integer and a single letter, \eg\ 3g. (These specify the node
and element identifier of the antenna within the ATA architecture.) An
antpol is specified by suffixing the antenna name with the letter X or
Y, \eg\ 2kY. We refer to a pair of antpols as a ``basepol.''

Although the ATA correlator produces cross-hand polarization data, the
calibration and analysis of these data have not yet been investigated
and tested at the time of this work. In this paper, cross-hand data
were discarded, and all sources were assumed to be completely
unpolarized. All images produced were of the Stokes I
parameter. Because the linear polarization of 3C~286 is $\sim$10\%,
this treatment will lead to $\sim$1\% errors in some
measurements. This effect contributes to, but does not dominate, the
typical uncertainty in our measurements (see \S\ref{s:uncerts}).

\subsection{Flagging}
\label{s:flag}

The ATA does not yet have an online flagging system and our
observations were performed outside of the protected bands. Fairly
extensive manual editing of the data was performed. Several automatic
criteria were applied. The 100 channels on both ends of the
1024-channel spectral window were flagged because the ATA digital
filter aliases in large amounts of noise at the bandpass
edges. Shadowed antennas were flagged, taking the effective antenna
radius to be 7.5 m (somewhat more than the 6.1 m physical diameter of
the primaries).  Finally, visibilities with $\sqrt{u^2+v^2} <
100\lambda$ were flagged as these short baselines were consistently
sensitive to terrestrial and solar RFI, undesired large-scale
structure, and possible cross-talk.

A fraction of the antpols had problems with their cryogenics, LNAs, or
delay calibrations, resulting in data for the night that could easily
be rejected after a cursory look.  Occasional correlator hardware
failures resulted in wildly erroneous amplitudes in a small subset of
the spectral data, \eg\ the first quarter of the spectrum of all
visibilities involving antpols 1aX, 1cY, and 4kX. Such data were also
flagged and the issue reported.

\begin{figure}[p]
\plotone{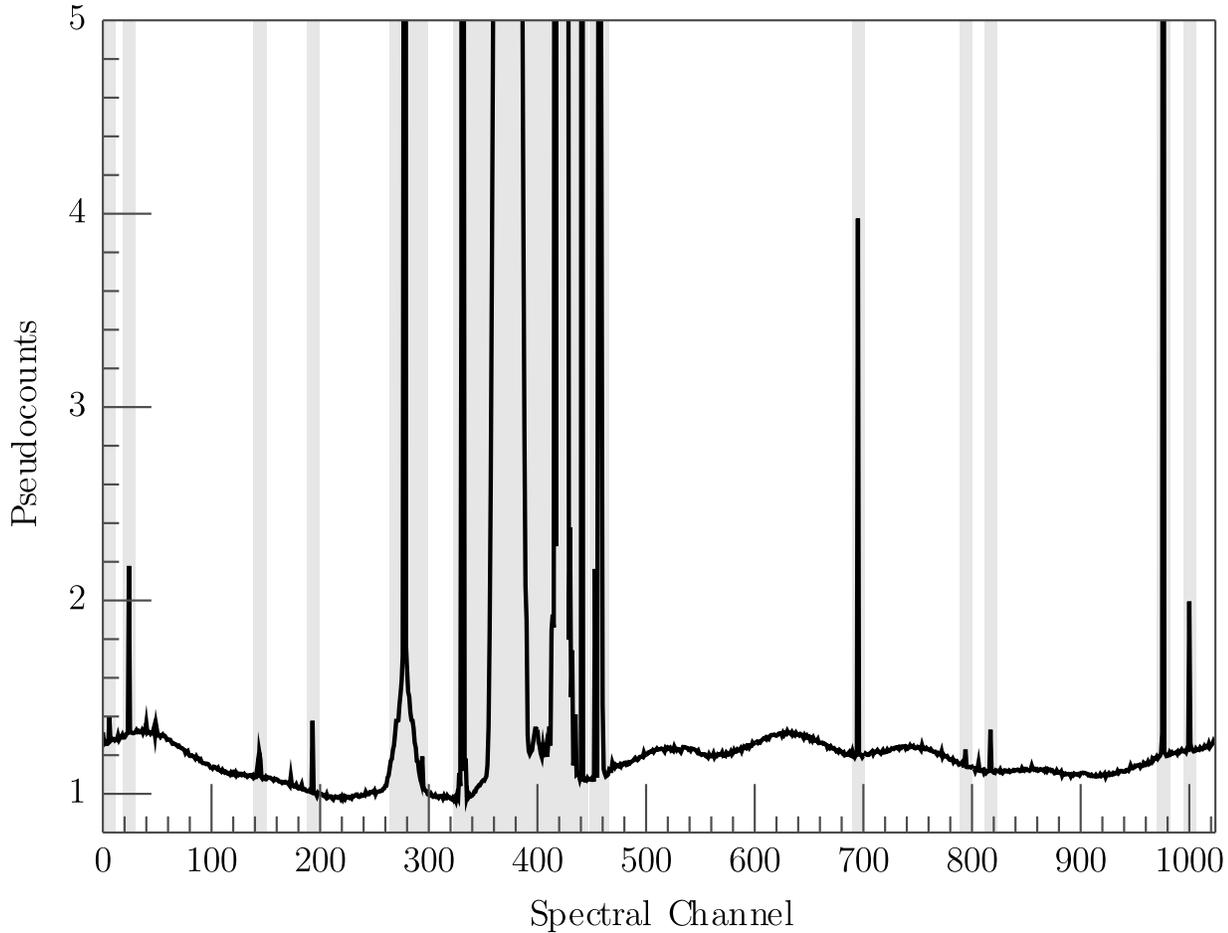}
\caption{An assessment of the RFI present during a block of 1.7 GHz
  observations. This plot is typical of those produced during
  execution of the semi-automated RFI excision algorithm described in
  \S\ref{s:flag}. {\it Solid lines} show summed amplitudes of the 1.7
  GHz data obtained during segment \#14 as a function of correlator
  channel, corresponding to frequencies ranging from 1.65 to 1.75 GHz.
  {\it Shaded regions} indicate RFI-afflicted regions detected by the
  peak-finding algorithm. Plots such as these were reviewed and
  sometimes edited to improve the flagging (\eg, to widen the flagged
  region around the broad peak centered near channel 280). Channel 512
  corresponds to a sky frequency of 1.700 GHz and the channel width is
  102.4 kHz. The sinusoidal features across the bandpass are due to
  the ATA digital filter.}
\label{f:rfiexamp}
\end{figure}

\begin{figure}[p]
\plotone{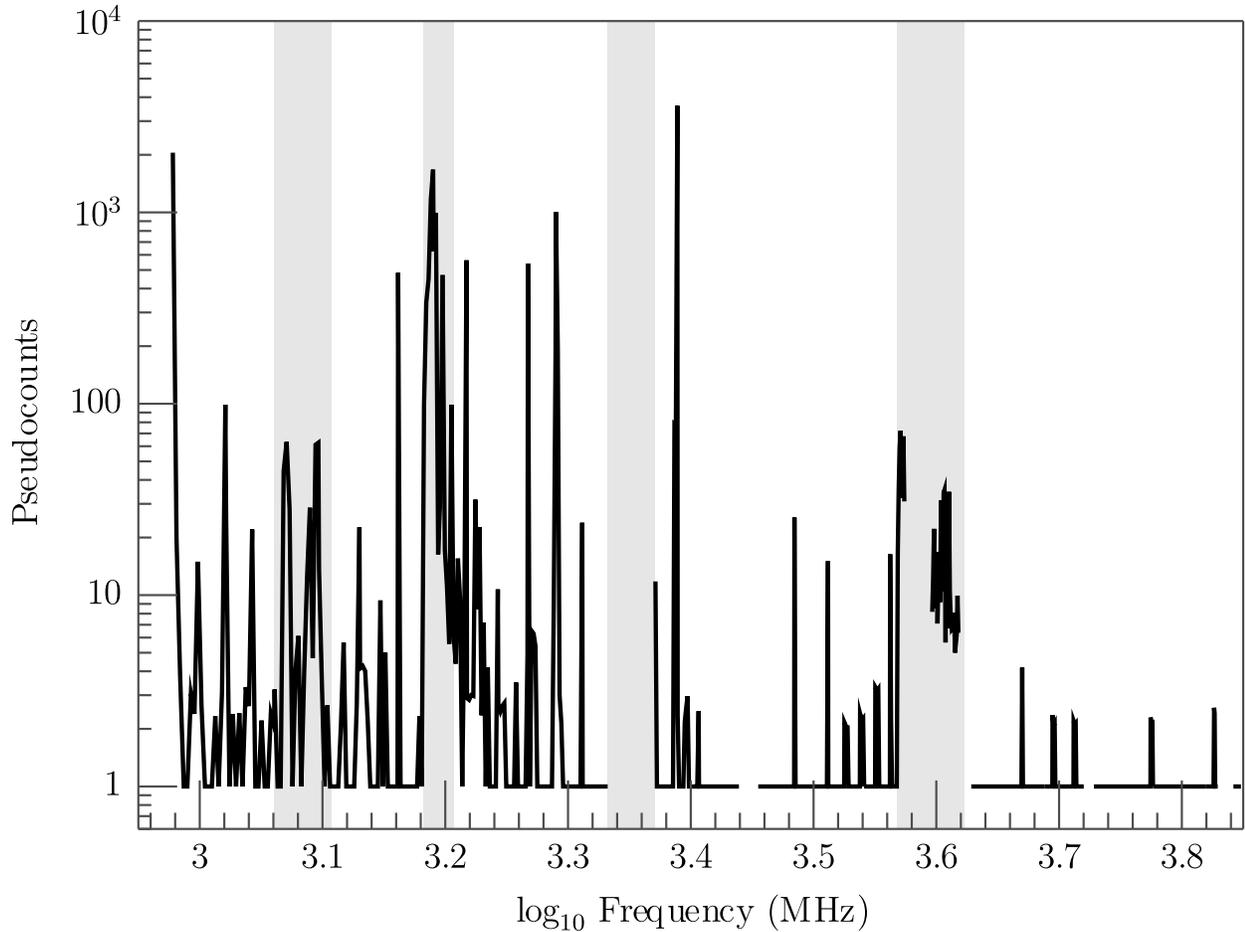}
\caption{The RFI environment of the ATA, in schematic form. {\it Solid
    lines} show a processed version of the summed data amplitudes, as
  described in \S\ref{s:flag}. The raw 100 kHz channelized amplitudes
  were processed to increase clarity while accurately portraying
  narrow RFI peaks. Gaps in the data are due to a lack of
  observations, usually because the RFI was known to dominate
  astronomical signals at certain frequencies. The {\it shaded
    regions} indicate bands afflicted with especially strong and
  persistent RFI.}
\label{f:rfi}
\end{figure}

After an initial check for bad antpols or correlator hardware, RFI was
flagged with a semi-automated, channel-based routine. {\em All} of the
amplitude spectra at a given frequency observed over the course of a
single segment were summed, regardless of time, baseline,
polarization, or pointing. A simple peak-finding routine identified
suspected RFI channels by searching for extremely high amplitudes. The
suspect channels were presented to the user graphically and made
available for interactive editing before a list of channels to flag
was recorded. The listed channels for each frequency were flagged out
for all observations at that frequency in the segment.
Figure~\ref{f:rfiexamp} shows an example of the summing and
peaking-finding process. Figure~\ref{f:rfi} characterizes the RFI
environment of Hat Creek via a schematic representation of the summed
amplitudes across the entire observed frequency range.

This procedure is obviously suboptimal in the case of time-varying
RFI, but it was found that most of the interference was virtually
always present. Informal tests of more sensitive methodologies that
attempted to better mitigate time-variable RFI failed to improve
results and were much more time-intensive to the user.

A few bands had nearly omnipresent RFI that made astronomical
observations nearly impossible. These were: 500-1000 MHz (terrestrial
television, radar, pagers, aeronautical navigation,
self-interference), portions of L band (1150-1280 MHz, aeronautical
navigation and GPS L5; 1520-1610 MHz, which includes GPS L1 at 1575
MHz), the lower end of S band (satellite communications, 2150-2350
MHz, especially Sirius XM radio centered at 2339 MHz), and C band
(satellite downlinks, 3700-4200 MHz).

After RFI-afflicted channels were flagged, various manual
data-quality checks were run to check for remaining issues. These
included phase-closure rankings of the calibrator observations,
evaluation of the amplitude and phase variability of the baselines,
and by-eye assessments of spectra. Many of these checks resulted in a
few antpols or basepols being flagged for the duration of the segment,
while occasionally a temporary system glitch was revealed and edited
out.

\begin{figure}[p]
\plotone{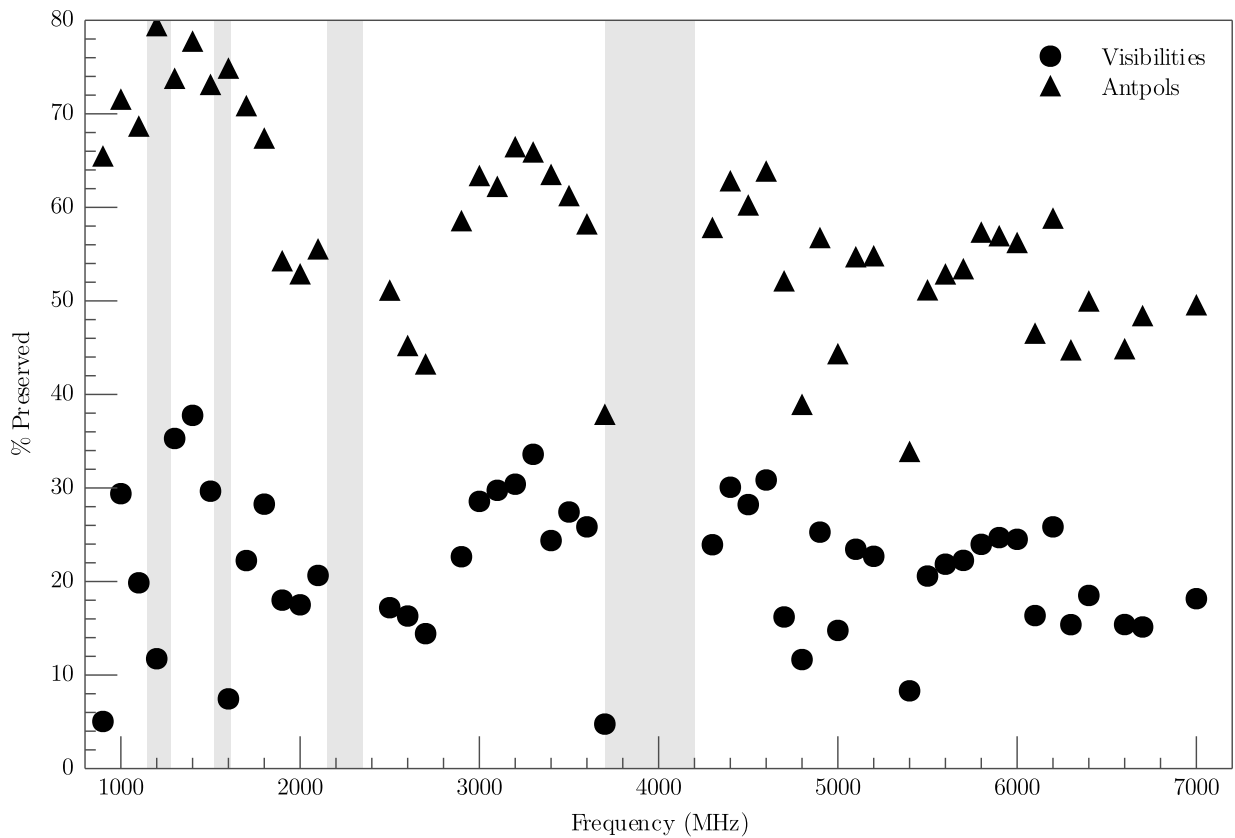}
\caption{Average data retention rates over all observations, as a
  function of observing frequency. Entirely discarded datasets are not
  shown. {\it Triangles} denote the fraction of antpols used during imaging
  as compared to the number present in the raw data. {\it Circles} denote
  the same for individual visibilities. The {\it shaded regions}
  represent bands afflicted with strong, persistent RFI. The maximum
  possible visibility retention rate is $\sim$75\% due to the flagging
  of edge channels, shadowed antennas, and short baselines described
  in \S\ref{s:flag}.}
\label{f:dataloss}
\end{figure}

Finally, after system temperatures were computed (see below), antpols
with a \tsys\ above 300 K were flagged. At the end of the flagging
process, $\sim$40\% of the antpols and $\sim$75\% of the visibility
channels in a typical dataset were discarded.  Figure~\ref{f:dataloss}
shows the average data retention rates as a function of frequency. The
deleterious effect of RFI is apparent, as is the moderate degradation
of system performance with increasing frequency.  Approximately 30\%
fewer antpols and visibilities are retained at 7 GHz as compared to
1.4 GHz.

Ongoing retrofits to the array are increasing the amount of usable
data generated during each observation. A large portion of this work
comprises improved RFI mitigation techniques, including better
shielding, especially of the signal processing room; improved
techniques and algorithms for interference excision; and stricter
on-site RFI protocols. Some data are lost because certain antpols have
extremely poor system temperatures, for various reasons: failed
cryogenics systems, damaged LNAs, or offset pointing due to imbalanced
electrical connections. More robust replacement hardware will make
these issues much rarer.

\subsection{Calibration}
\label{s:calimg}

The calibration of our data is challenging because of the modest \uv\
coverage, short integration times, and wide temporal and spatial
separation between snapshots entailed by our observing mode. We will
provide an overview of our calibration process with more detailed
descriptions presented in the subsections below.

The snapshot observations of the calibrator sources were first
self-calibrated to point-source models. All observations of the same
calibrator and frequency were then imaged jointly to form a
self-calibrated {\it reference image} with \uv\ coverage and
sensitivity superior to those obtained in any one snapshot. The
individual calibrator snapshots were then recalibrated to their
corresponding reference images.

The snapshot observations of the science sources then had
appropriate calibration solutions applied to them. Some snapshots (of
both science targets and calibrators) were duplicated and had several
different calibration solutions applied to allow for cross-checks of
the calibrator flux models. Reference images were then generated from
the calibrated observations and once again the individual snapshots
were recalibrated to their corresponding reference images.

Given the nature of this work, accurate broadband calibrator flux
density models are vital. The 1999.2 analytic models listed in the VLA
Calibrator Manual were initially adopted and found to yield good
results. These models were, however, almost a decade old at the time
of this work, and long-term VLA monitoring indicates that the fluxes
of the calibrators evolve slightly on this timescale. Section
\ref{s:modcheck} describes consistency checks used to verify and
fine-tune the flux density scale used in this work.

\subsubsection{Self-calibration}
\label{s:selfcal}

The calibration process started with self-calibration of the
calibrator observations. Bandpasses and complex antenna gains in the
individual snapshots were first determined via the MIRIAD task {\tt
  smamfcal}, which assumes a point-source model. Spectral smoothing
was used to increase the signal-to-noise ratio (SNR) in the
high-frequency observations.

System temperature (\tsys) information was then computed and inserted
into the datasets. This step was necessary because the ATA did not
have online system temperature calibration at the time of this work. A
Python script called {\tt calctsys} was developed to estimate \tsys\
values from the variances in the real and imaginary parts of the
visibilities across the band. The variances were used to compute
per-basepol \tsys\ values assuming an antenna gain of 153 Jy/K.
Per-antpol \tsys\ values were then computed by performing a
least-squares minimization assuming that the \tsys\ of each basepol
was the geometric mean of the \tsys\ of its two contributing antpols.
The values obtained in this way had good reproducibility over the
course of a segment and agreed well with lunar measurements made by
ATA staff (R. Forster, private communication). The system temperatures
could vary strongly as a function of frequency, with most, but not
all, antpols demonstrating superior performance at lower frequencies.

\begin{figure}[p]
\centering
\leavevmode
\includegraphics[width=0.517\columnwidth]{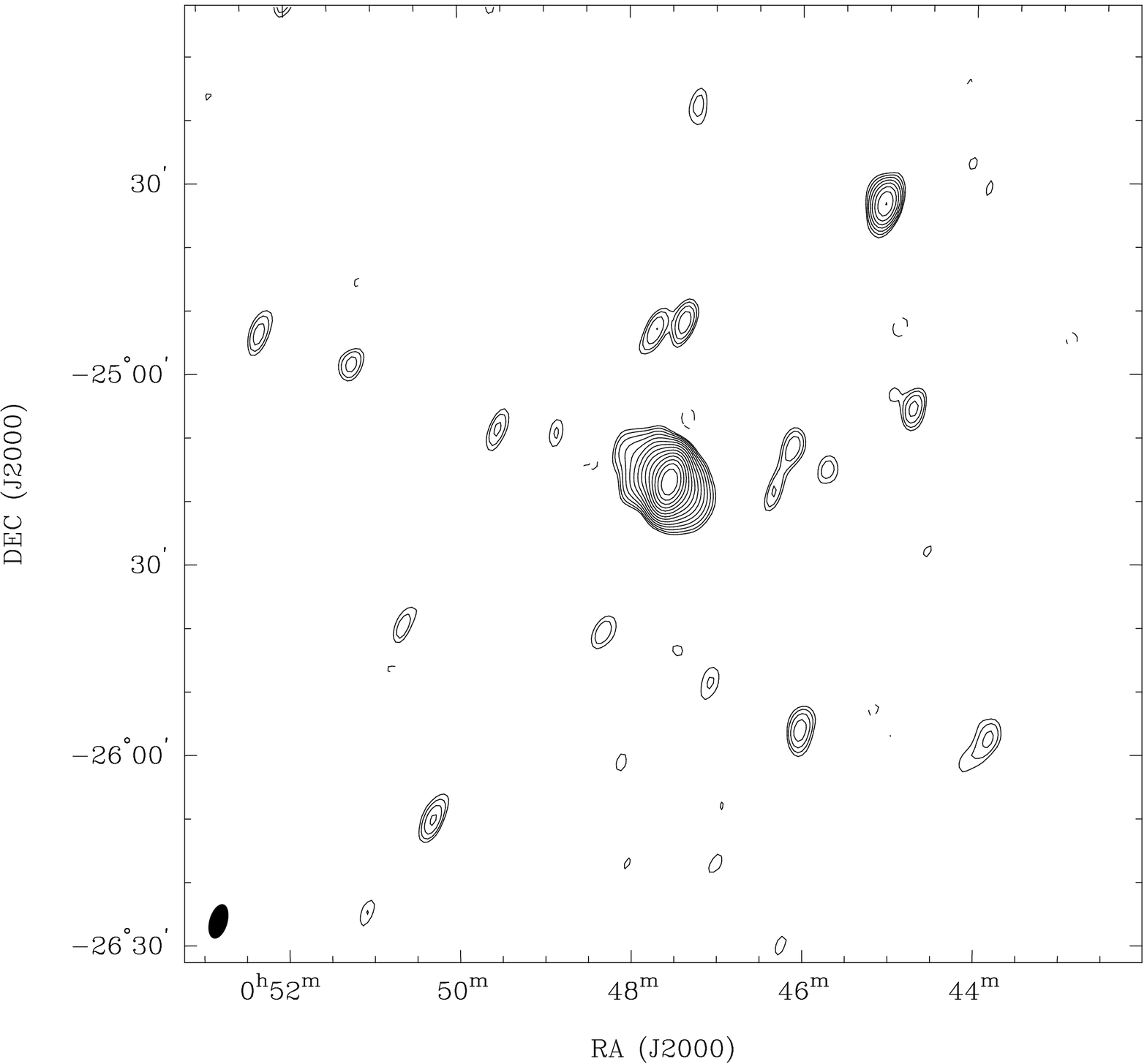}%
\hfil
\includegraphics[width=0.433\columnwidth]{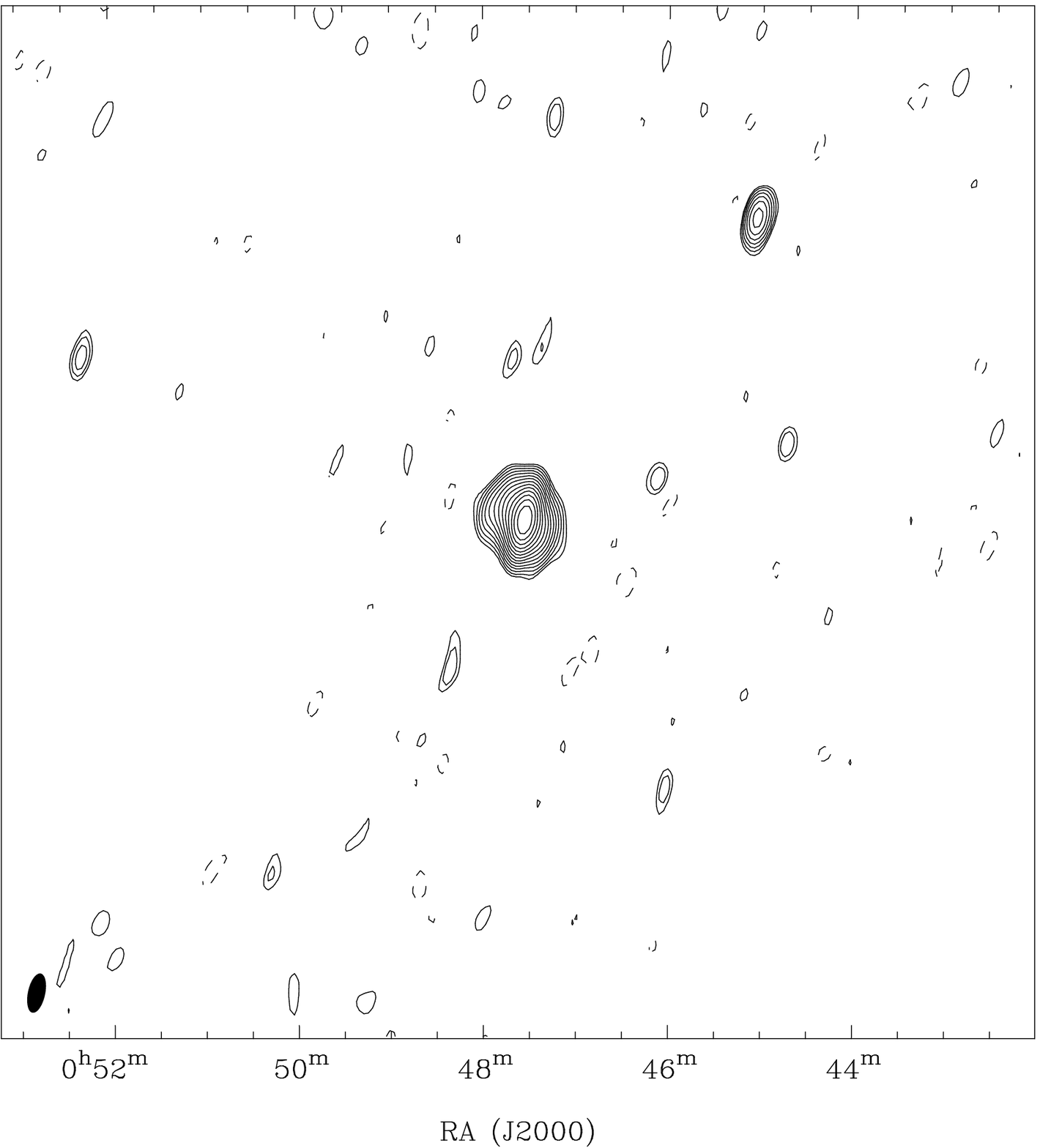}%
\caption{ATA maps of NGC~253 at 1.3 GHz. The {\it left panel} shows
  the reference image for this source and frequency while the {\it
    right panel} shows an image generated from a 70~s integration made
  during Segment \#3. The total integration time for the reference
  image is 250~s. Contour levels in both panels are $30 \textrm{ mJy}
  \times 2^{n/2}$ for $n = 0, 1, 2, \ldots, 13$. The RMS noise levels
  of the reference image and the snapshot are 9 and 11 mJy,
  respectively.  Although the longer integration time and more
  complete \uv\ coverage of the reference image lead to a noticeable
  increase in its quality, the total fluxes of NGC~253 measured in the
  two images agree to 2\%.}
\label{f:refsnap}
\end{figure}

\begin{figure}[p]
\plotone{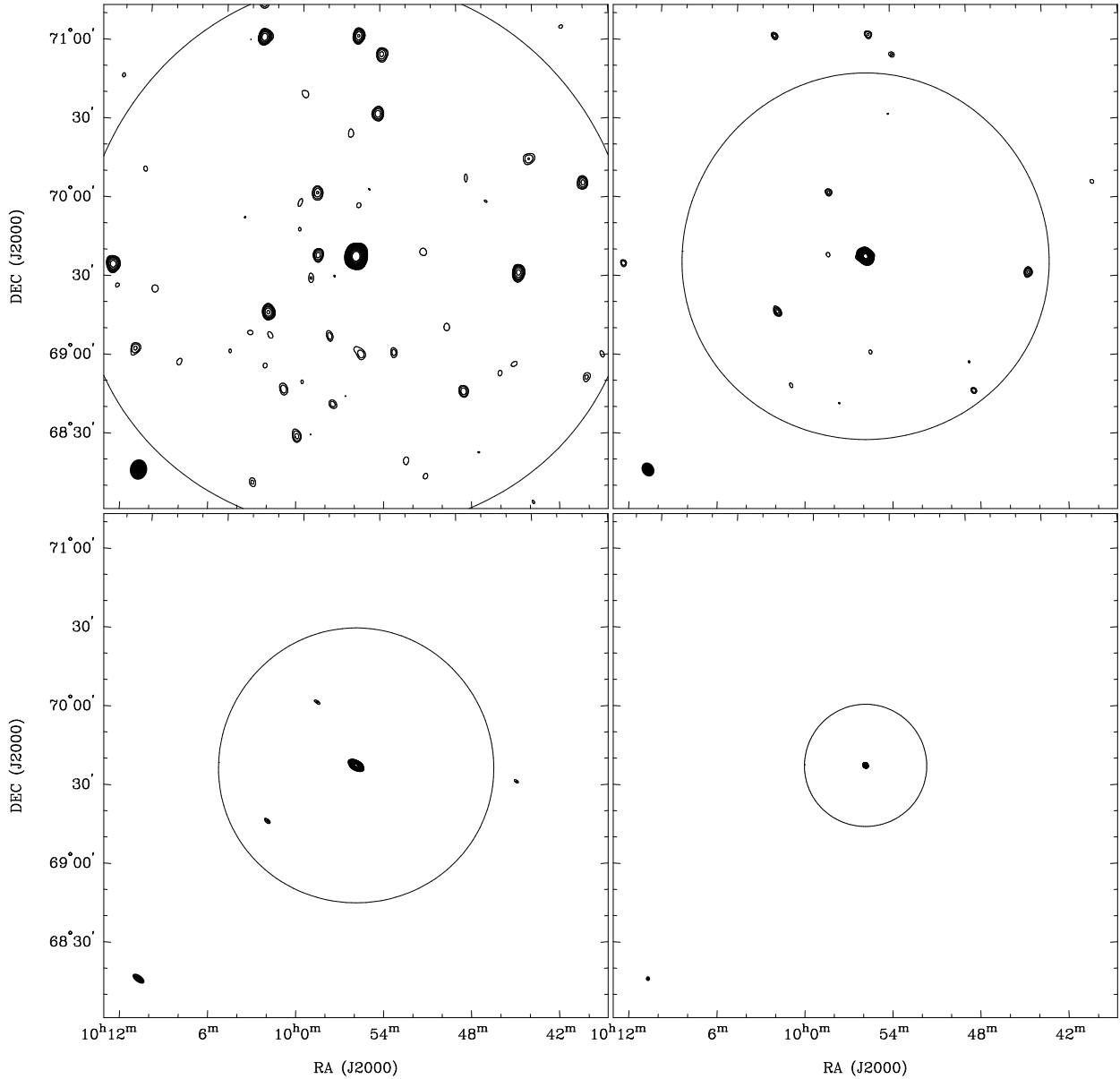}
\caption{ATA maps of M82 at 1.0, 1.5, 2.0, and 4.5 GHz ({\it top
    left}, {\it top right}, {\it bottom left}, and {\it bottom right}
  panels, respectively). The depicted maps are the reference images at
  each frequency reimaged onto a common coordinate grid. {\it
    Circles:} half-power points of the ATA beam. {\it Filled
    ellipses:} synthesized beamsizes. Contour levels in all panels are
  $50 \textrm{ mJy} \times 2^{n/2}$ for $n = 0, 1, 2, \ldots, 13$. The
  RMS noise levels of the images are 17, 13, 10, and 11 mJy. The
  number of sources seen decreases as frequency increases due to both
  the shrinking primary beam of the telescope and the general downward
  slope of the radio spectra of celestial sources.}
\label{f:mfref}
\end{figure}

Reference images were then created from the group of calibrator
snapshot observations. All of the snapshots of each source and
frequency were merged into a temporary dataset that was imaged using
natural weighting. Multifrequency synthesis \citep{ccw90} was used to
avoid bandwidth smearing and provide better \uv\ sampling. The images
were deconvolved with the MIRIAD implementation of the CLEAN algorithm
\citep{h74}, which incorporates the algorithmic adjustments suggested
by \citet{c80} and \citet{sdi84}. After an initial round of imaging,
the merged visibilities were self-calibrated to the generated CLEAN
component model and imaged once more. The second self-calibration was
phase-only if the SNR of the data was less than 0.3. Testing showed
that further iterations of calibration and imaging did not lead to
improvements in the image quality.  Figure~\ref{f:refsnap} shows the
reference image of NGC~253 at 1.3 GHz and an image generated from a
single snapshot observation of the same source at the same frequency.
Figure~\ref{f:mfref} shows reference images of M82 at four
frequencies. The individual calibrator snapshots were then
recalibrated using the CLEAN component model of the reference images.

\subsubsection{Secondary calibration}
\label{s:copycal}

The complex antenna gains and bandpass solutions were copied from each
calibrator snapshot to any observations made at the same frequency
within 40 minutes of the calibrator observation, including
observations of different calibrators. To allow further consistency
checks, if multiple calibrator observations could be applied to a
given snapshot, it was duplicated and a different calibration was
applied to each copy. The copy with the smallest time differential
between the two observations was treated as the ``official'' version
of the observation, with the other copies not being included in the
final analysis. The median time differential of
the ``official'' observations was 21 minutes, with values ranging from
3 to 38 minutes. The stability of the ATA over such timescales was
qualitatively very good. Such stability is particularly noteworthy
because of the large angles on the sky separating the calibrator and
target sources (see, \eg, Table~\ref{t:oscans}).

Once the observations of the science targets were flux and bandpass
calibrated, system temperatures were computed and inserted into the
datasets using {\tt calctsys}. If the science targets had bright,
off-center emission, the {\tt calctsys} technique would have computed
incorrect \tsys\ values due to significant visibility variations
across the passband, but this was not an issue for our selected
sources.

Reference images of the science targets were then generated from the
calibrated observations in a manner similar to that described above.
A reference image was generated for each observed combination of
source, sky frequency, and calibrator. Before imaging, the merged data
were phase self-calibrated against a primary-beam-corrected model
generated from the NRAO VLA Sky Survey (NVSS; \citet{c+98}) catalog.
The purpose of this step was to compensate for occasional
antenna-based phase glitches that could occur between the calibrator
observation and the observation it was applied to. To roughly correct
for the fact that the NVSS data only comprise a single frequency, each
NVSS model image was scaled by $S_\nu(\nu)/S_\nu(1.4\textrm{ GHz})$
before the phase self-calibration, where $\nu$ was sky frequency of
the ATA observations and $S_\nu(\nu)$ was a simplistic model of the
spectrum of the source being observed. The NVSS catalog sources
corresponding to NGC~253 was found to poorly reflect its extended
structure, so a bootstrapped model of that source's spatial emission
was used rather than the NVSS model. After this first round of phase
self-calibration, the \uv\ data were imaged and recalibrated as
described above.

\subsection{\uv\ modeling}
\label{s:uvmodel}

Fluxes were determined by fitting the \uv\ data with point-source and
elliptical Gaussian models based on the reference images. Our
measurements are listed in Tables~\ref{t:m82}--\ref{t:apg220}. The
calibrators and Arp~220 were all well-modeled as point sources. M82
was modeled as an elliptical Gaussian. The deconvolved source sizes
determined for M82 varied slightly over the range of frequencies
observed. 
The major (minor) axis of M82 measured approximately 60 (40) arcsec at
1 GHz and 35 (20) arcsec at 7 GHz, with a roughly constant position
angle (PA) of $\sim$80$\,^\circ$ (east from north). The
higher-frequency values are consistent with observations made at 1.48,
4.87, 32, and 92~GHz \citep{sbb85,kwm88,ck91}. The lower-frequency
values are likely biased by the large ($\sim$250\arcsec) synthesized
beam of the ATA at low frequencies.

NGC~253 was modeled as the superposition of an elliptical Gaussian and
a more compact source. At lower frequencies, the field of view
contained bright sources away from the pointing center which were also
modeled. The size of the extended emission is strongly
frequency-dependent, having a major (minor) axis of 480 (200) arcsec
at 1 GHz and 200 (50) arcsec at 7 GHz, with roughly constant PA of
$\sim$50$\,^\circ$. The morphology of this region at 1.5~GHz is
consistent with the ``inner disk'' identified in \citet{hsh84}. One
should bear in mind that the spectra presented in Table~\ref{t:ngc253}
and Figure~\ref{f:s253_500} manifest not only the changing brightness
of this region but its changing size as well. The more compact
source was unresolved at lower frequencies and modeled as a point
source. Above 3.5 GHz, however, it was possible for some structure in
the compact source to be discerned, and it was found that an
elliptical Gaussian with major (minor) axes of $\sim$16 ($\sim$10)
arcsec at a PA of $\sim$60$\,^\circ$ was a practical model. This
morphology is consistent with the ``nuclear peak'' seen in the near IR
by \citet{s+85}.

The residuals to the fitted \uv\ models were imaged to check for
missed sources, poorly calibrated baselines, and so on. Further
flagging was usually sufficient to solve any issues. Some datasets
with severe RFI were not recoverable and had to be discarded.

\begin{figure}[p]
\plotone{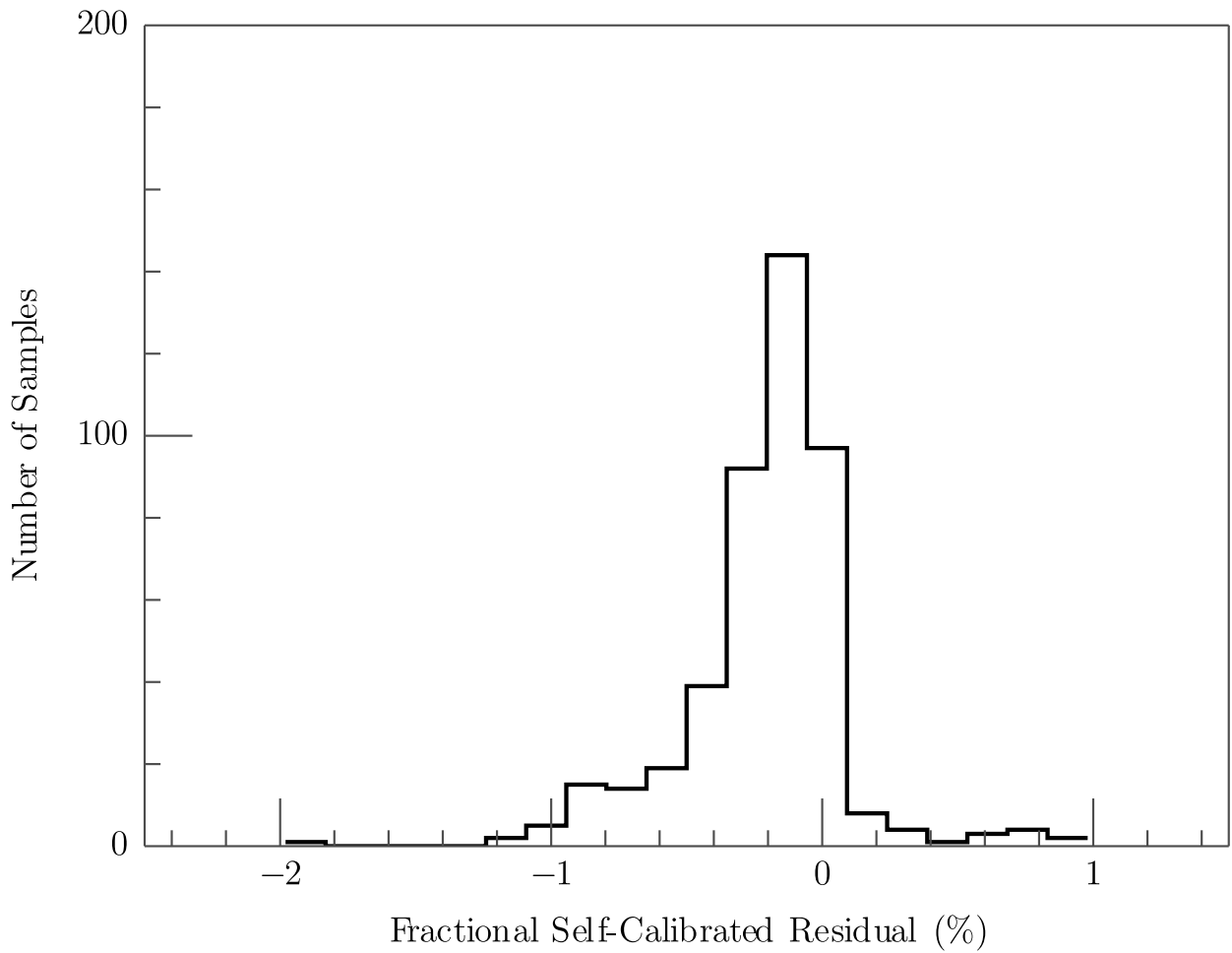}
\caption{Histogram of fractional flux residuals between calibrator
  observations and analytic calibrator models. The fluxes for the
  calibrator observations are derived from \uv\ modeling of calibrator
  data that have been self-calibrated to the calibrator reference
  images.}
\label{f:sccheck}
\end{figure}

Figure~\ref{f:sccheck} shows a histogram of the fractional flux
residuals of the calibrator snapshots as compared to the analytic
calibrator models. The calibrator fluxes are those derived from \uv\
modeling of calibrator snapshots that have been self-calibrated to the
calibrator reference images. The vast majority of the measurements
agree to within 1\%, demonstrating that the reference-image
calibration pipeline propagates fluxes with high fidelity.

\subsection{Uncertainties}
\label{s:uncerts}

The MIRIAD \uv\ fitting task, {\tt uvfit}, reports statistical
uncertainties, but these fail to account for systematics and are only
lower limits. We augmented these uncertainties with both relative
and absolute terms:
$$
\sigma_i^2 = s_i^2 + (\eta S_{\nu,i})^2 + (\sigma_a \cdot
1\textrm{ Jy})^2,
$$
where $\sigma_i$ is the uncertainty used for the $i^\mathrm{th}$
measurement, $s_i$ is the uncertainty for that measurement
reported by the fitting routine, $S_{\nu,i}$ is the value of the
measurement, and $\eta$ and $\sigma_a$ are tunable parameters.

The choice of these parameters was derived from the statistics of the
ensemble of observations. For each observed combination
of source, frequency, and calibrator, the weighted mean of the
contributing observations was computed:
$$
\bar S_\nu = \sum_{i=1}^n S_{\nu,i} \sigma_i^{-2} /
\sum_{i=1}^n \sigma_i^{-2}.
$$
Let there be $N$ observed combinations of source, frequency, and
calibrator, with $n_j$ observations of the $j^\mathrm{th}$
combination. Let $\bar S_{\nu,j}$ be the weighted mean of the $n_j$
observations, and $S_{\nu,j,i}$ and $\sigma_{j,i}$ the values and
uncertainties of the individual observations.  If the uncertainties
have been correctly assessed and the measurements are drawn from
normal distributions, the sum of the squares of the normalized
residuals,
$$
K^2 = \sum_{j=1}^N \sum_{i=1}^{n_j} \left(\frac{S_{\nu,j,i} - \bar
    S_{\nu,j}}{\sigma_{j,i}}\right)^2,
$$
should be taken from a $\chi^2_k$ distribution, where $k =
\sum_{j=1}^N n_j - N$ is the number of degrees of freedom. The
expectation value of such a distribution is $k$.

\begin{figure}[p]
\plotone{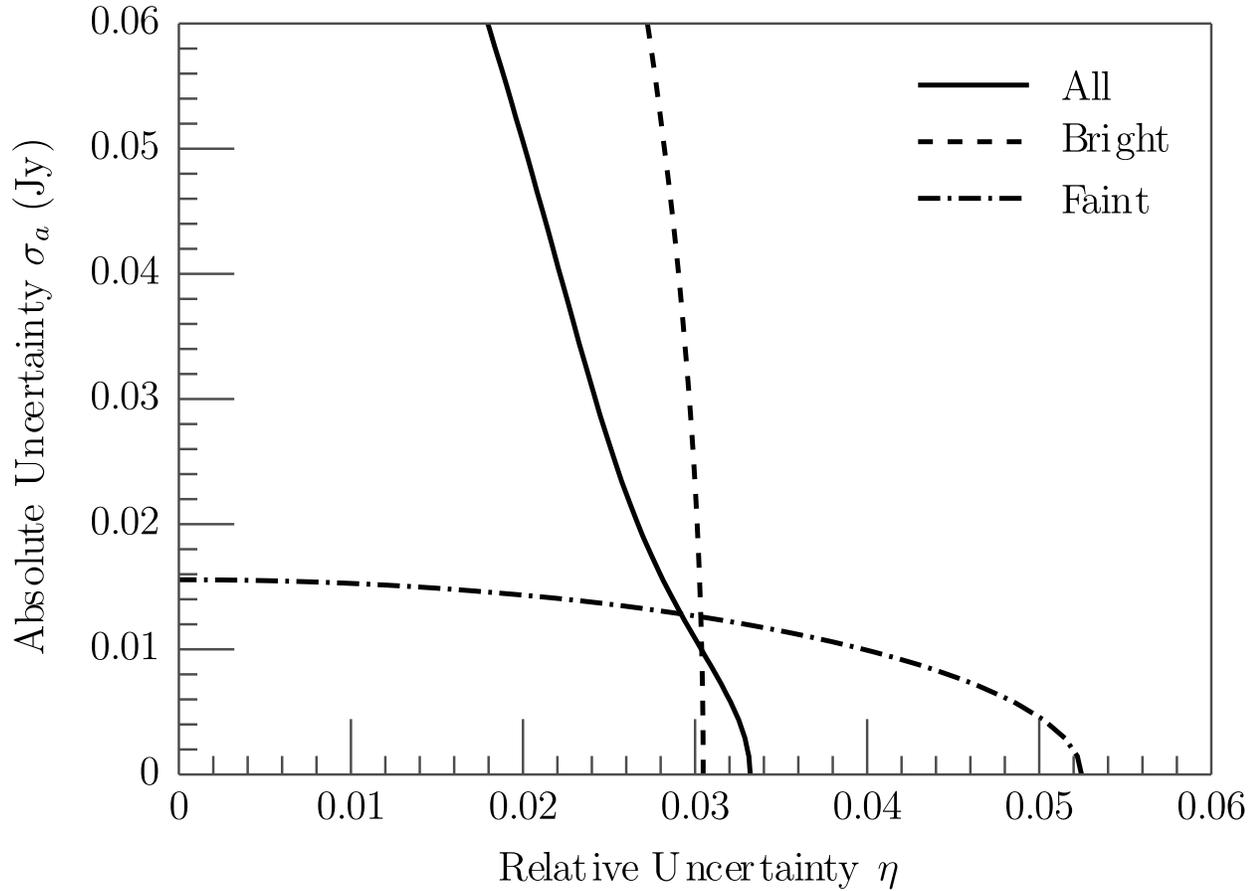}
\caption{Contours of $K^2 = k$ in the $\eta$-$\sigma_a$ space of
  uncertainty augmentation parameters. See \S\ref{s:uncerts} for
  discussion. {\it Solid line:} the contour for $K^2$ values derived
  from all repeated flux measurements. {\it Dashed line:} the contour
  for values derived from measurements of sources with $S_\nu > 3$ Jy.
  {\it Dot-dashed line:} the contour for values derived from
  measurements of sources with $S_\nu < 1$ Jy.}
\label{f:ucheck}
\end{figure}

The value of $K^2$ was computed on a grid of $\eta$ and $\sigma_a$ for
three groups of measurements: the full ensemble, those with $\bar
S_\nu < 1$ Jy (``faint''), and those with $\bar S_\nu > 3$ Jy
(``bright''). Figure~\ref{f:ucheck} shows contours of $K^2 = k$ (\ie,
uncertainties agreeing with observed scatter) in the $\eta$-$\sigma_a$
plane. As expected, the contours of constant $K^2$ are primarily along
the $\eta$ axis for the faint group (for which $\sigma_a$ dominates
the additional uncertainty) and primarily along the $\sigma_a$ axis
for the bright group (for which $\eta$ dominates the additional
uncertainty). The $K^2 = k$ contours for the three groups intersect at
an approximate location of $\eta = 0.030$ and $\sigma_a = 0.014$, \ie\
an additional 3.0\% fractional uncertainty and 14 mJy absolute
uncertainty. These are the values of $\eta$ and $\sigma_a$ that were
adopted. The value of $\sigma_a$ is approximately equal to the thermal
noise in good conditions and is about half of the thermal noise at
high frequencies, where \tsys\ generally increases.

If the uncertainties are correctly assessed and measurements are drawn
from normal distributions, the sum of the normalized residuals of each
measurement is drawn from a normal distribution with mean 0 and
variance $\sigma^2 = \sum_{j=1}^N n_j$. For the chosen parameter
values, the sum of the residuals was $0.52 \sigma$. 
This moderate value suggests that the adopted uncertainties
characterize well the observed scatter in the measurements and that
the weighted-mean approach does not mask large variations
in the individual measurements.

\subsection{Spectra}
\label{s:spectra}

\begin{figure}[p]
\plotone{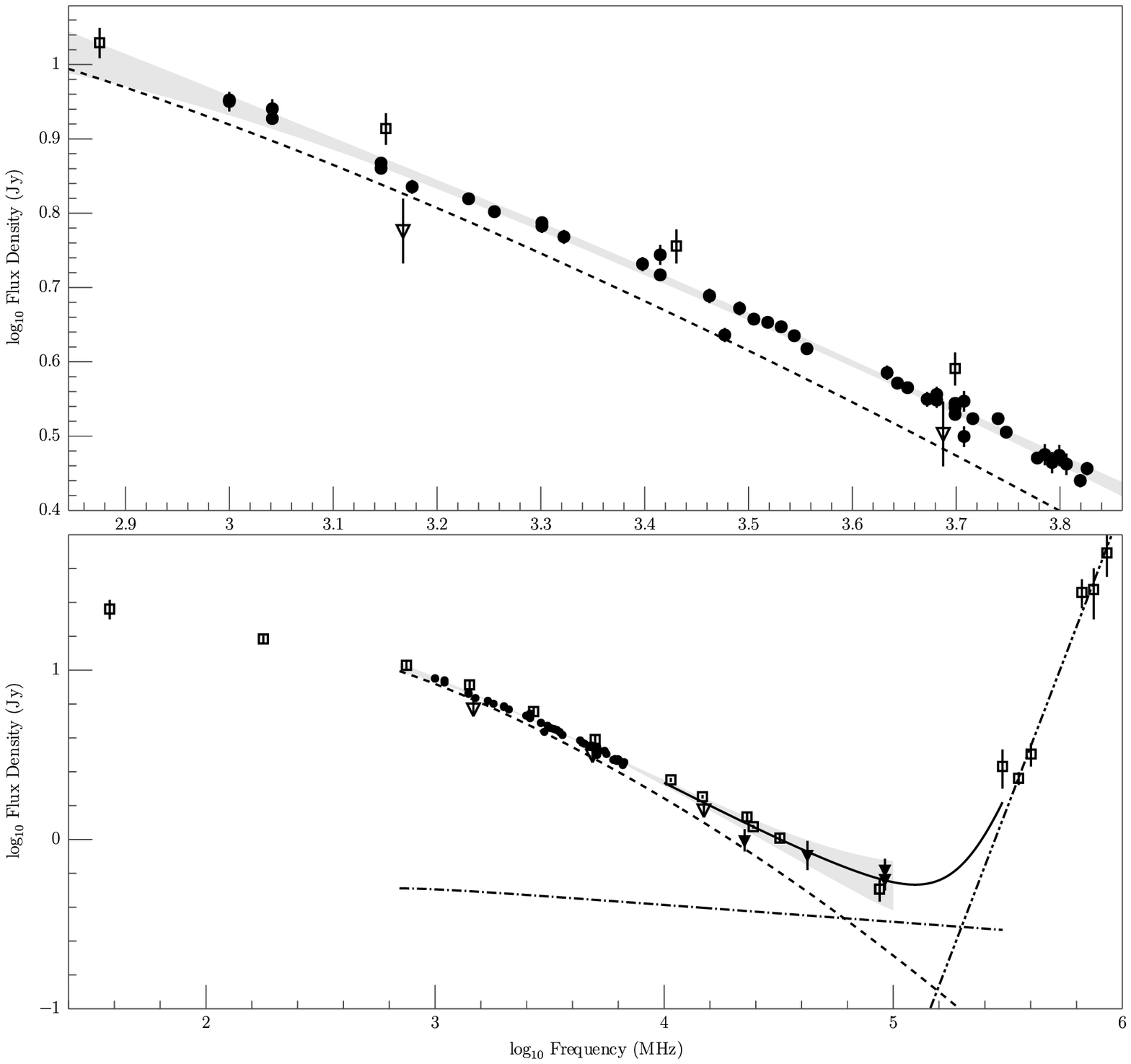}
\caption{Spectrum of M82 with model. The two panels show different
  views of the same data. {\it Circles:} new measurements presented in
  this paper. Other symbols indicate data from the literature:
  \citet{ck91} and citations therein; \citet{kwm88} and citations
  therein; \citet{jhm78,kekw79,lr09,rvzga04,scbb96}.  These
  measurements are made with various telescopes in various
  configurations and, due to the properties of interferometers, are
  not always directly comparable. {\it Squares:} data obtained with
  single dishes. {\it Triangles:} data obtained with interferometers.
  {\it Filled symbols:} data used in the fits described in
  \S\ref{s:sf82}. {\it Open symbols:} data not included in the fits.
  {\it Shaded region:} the 3$\sigma$ uncertainty region of the model
  spectrum as described in \S\ref{s:modres}. {\it Dashed line:} the
  model nonthermal contribution. {\it Dash-dotted line:} the model
  thermal contribution. {\it Dash-double-dotted line:} a power-law fit
  to the FIR data. {\it Solid line:} the sum of the thermal,
  nonthermal, and FIR power law contributions. See \S\ref{s:mr82} for
  further discussion of the model results.}
\label{f:s82}
\end{figure}

We constructed broadband, high-resolution spectra out of the flux
measurements described in \S\ref{s:uvmodel}. Repeated measurements
were combined by taking a weighted mean as described in the previous
section.
Twenty-five percent of the spectral data are based on a single
observation, while 63\% are based on at least three observations. The
maximal number of contributing observations is 13, of M82 at 5.0~GHz.
The uncertainty of each spectral point is the propagated uncertainty
of the weighted mean,
$$
\sigma^2_{S,j} = \left(\sum_{i=1}^{n_j} \sigma_{j,i}^{-2}\right)^{-1}.
$$
The spectra are shown in Figures \ref{f:s82}-\ref{f:s220} along with
measurements from the literature and model results as discussed in
\S\ref{s:model}.

\subsection{Self-consistency Checks}
\label{s:modcheck}

We performed no absolute flux calibration of the ATA. The measurements
we have made are thus measurements of flux density ratios, with
absolute flux densities being derived from {\it a priori} spectral
models of the calibrator sources. We cannot verify the models
themselves, but we can assess whether our measured flux ratios are
consistent with them.

Let the ``true'' flux density of source $X$ (\ie, the value
that would be measured by a perfect, absolutely calibrated antenna at
the time of observation) be $T_X$, the {\it a priori} modeled flux
density of $X$ be $M_X$, and the flux density of $X$ observed by the
ATA after calibration to source $Y$ be $O_{X,Y}$. All of these values
vary with frequency. In the simplest possible model, at a fixed
frequency $O_{X,Y}$ and $T_X$ are directly related by a
calibrator-dependent gain factor $g_Y$:
$$
O_{X,Y} = g_Y T_X.
$$
The gain is self-calibrated such that $O_{Y,Y} \equiv M_Y,$ so that
ideally
$$
O_{X,Y} = M_Y \frac{T_X}{T_Y} \equiv M_Y R^T_{X,Y},
$$
where $R^T_{X,Y}$ is the true flux density ratio of $X$ to $Y$. Not
knowing whether this simple model is correct, we will define the
observed flux ratio as
$$
R^O_{X,Y} \equiv \frac{O_{X,Y}}{M_Y}.
$$
If this simple calibration model holds, the observed ratios should
demonstrate consistency and closure:
\begin{gather}
\label{e:calcheck1}
R^O_{X,Y} \cdot R^O_{Y,X} = 1, \\
\label{e:calcheck2}
R^O_{X,Y} \cdot R^O_{Y,Z} \cdot R^O_{Z,X} = 1.
\end{gather}
Furthermore, if the spectral models are accurate, it should be found
that
\begin{equation}
R^O_{X,Y} = R^M_{X,Y} \equiv M_X / M_Y.
\end{equation}
(If $M_X$ is not known precisely, two measurements $R^O_{X,Y}$ and
$R^O_{X,Z}$ can still be used to check the model ratio $R^M_{Y,Z}$ by
computing $R^O_{Y,Z} = R^O_{X,Z} / R^O_{X,Y}$.) If all three of these
properties are observed, there is good reason to believe that
$R^O_{X,Y} = R^T_{X,Y}$ and $O_{X,Y} = T_X$, though it is impossible
to demonstrate the latter without absolute flux measurements.

We compute these ratios and plot them in Figure~\ref{f:consistency}.
Our measurements do not have enough precision for us to believably
solve for spectral models of the calibrators from scratch. We do,
however, solve for a change in the overall scaling of the VLA 1999.2
models, \ie, for three parameters $f_{48}$, $f_{147}$, and
$f_{286}$ such that $M_X = f_X V_X$, where $V_X$ is the VLA model of
source $X$ and we denote the calibrators by their 3C numbers. Given
the constraints implied by Equations~\ref{e:calcheck1} and
\ref{e:calcheck2}, there are only two independent $R^O_{X,Y}$, so a
normalizing assumption must be applied to derive the three $f_X$ from
observations. We choose to fix $f_{48} f_{147} f_{286} = 1$, which is
equivalent to requiring that the total relative change to the models,
$\ln^2 f_{48} + \ln^2 f_{147} + \ln^2 f_{286}$, be minimized.

To solve for the $f_X$ from the data, we define $a = f_{48} / f_{147}$
and $b = f_{48} / f_{286}$.
Minimizing the logarithms of $R^M_{X,Y} / R^O_{X,Y}$ in a
least-squares sense leads to $a = 1.018$ and $b = 1.007$ with a root
reduced $\chi^2$ parameter (\srcs) of 1.07.
This minimization included two precise measurements made at the VLA in
2008 September: $R^O_{48,286} = 1.066 \pm 0.003$ at 1.465 GHz and
$R^O_{48,286} = 0.7349 \pm 0.002$ at 4.885 GHz (R. Perley, private
communication). Using our normalizing assumption, we find $f_{48} =
1.011$, $f_{147} = 0.989$, and $f_{286} = 1.001$.  This is consistent
with evidence from the VLA that the flux density of 3C~48 has recently
risen slightly (R. Perley, private communication).

Our measured flux ratios are well-described within their uncertainties
by the rescaled VLA models with the exception of the $R^O_{147,286}$
data. These observations were made in June 2009, well after the main
observing for this work was completed, after it was realized that
there were only a few data comparing these calibrators. Unlike the
2008 observations, these occurred in the daytime, and solar
interference was a significant problem with the low-frequency data.
Nevertheless the high-frequency ratios, for which solar interference
was much less relevant, disagree significantly with the adopted
models. We have not been able to identify the cause of this
disagreement, but we do not believe that it is due either to genuine
inaccuracy in the models or to problems in the calibration and \uv\
modeling pipeline. If the former were the case, it would represent an
unprecedented variation in the flux of some combination of 3C~286 and
3C~147. If the latter were the case, the consistency property of
Equation~\ref{e:calcheck1} would not hold as well as it does in the
observations. We believe the robustness of the rest of the results
provide ample reason to believe that the models and reduction pipeline
are generally robust and accurate.

\section{Spectral Modeling}
\label{s:model}

In order to assess the steepening of the nonthermal components
of the observed spectra, we modeled them with a two-component model
similar to that used in \citet{nkw97}:
\begin{equation}
\label{e:smodel}
S_\nu = S_\mathrm{th}(\nu) + S_\mathrm{nt}(\nu),
\end{equation}
where $S_\mathrm{th}(\nu)$ is the thermal contribution and
$S_\mathrm{nt}(\nu)$ is the nonthermal contribution. Free-free
absorption (FFA) is likely to play a significant role in determining
the spectra of our target galaxies. We attempt to account for its
effects by including as a parameter in our model $\tau_1$, the
free-free optical depth at 1~GHz. The thermal component is then
\begin{equation}
S_\mathrm{th}(\nu) = B_\nu(T_e) \Omega (1 - e^{-\tff}),
\end{equation}
where $T_e \simeq 10^4$ K is the temperature of the thermal electrons,
$\Omega$ is the angular size of the emitting region, and \tff\ is the
free-free optical depth as a function of frequency. We use the
canonical frequency dependence of the free-free optical depth,
\begin{equation}
\tff = \tau_1 \nu_g^{-2.1},
\end{equation}
where $\nu / \mathrm{GHz} = \nu_g$. In typical cases, the large-scale
thermal emission in galaxies at cm wavelengths is optically thin
\citep[\eg][]{dbg88,pd92}, $\tff \ll 1$, in which case the thermal
emission has the well-known frequency dependence of $S_\mathrm{th}
\propto \nu^{-0.1}$. The galaxies we consider are, however, decidedly
atypical, with dense, compact emission regions, so we avoid this
approximation in our modeling. Our results indicate, however, that it
broadly holds even for the galaxies in our sample (cf. Figures
\ref{f:s82}-\ref{f:s220}).

For simplicity we assume that the thermal and nonthermal components
are well-mixed and hence share the same \tff. If this is not the case
\citep[as is true of the Milky Way, {\it e.g.}][]{br85}, this
assumption would require loosening in a more detailed analysis,
although at small values of \tff\ its impact is minimal. In the
well-mixed case, the nonthermal emission is attenuated so that
\begin{equation}
S_\mathrm{nt}(\tnu) = \tilde S_\mathrm{nt}(\tnu) \cdot \frac{1 - e^{-\tff}}{\tff},
\end{equation}
where $\tilde S_\mathrm{nt}$ is the unabsorbed nonthermal spectrum.
Because our fits model the thermal and nonthermal contributions
simultaneously and we do not approximate any FFA-related effects, we
should obtain good results even when $\tau_1$ is large. We model the
unabsorbed nonthermal component as
\begin{equation}
\label{e:nonth}
\tilde S_\mathrm{nt}(\tnu) = 10^{A + B \log_{10}\tnu + C
  \left[\log_{10}\tnu\right]^2} \mathrm{Jy},
\end{equation}
a parabola in log-log space. Our parabolic spectral model is not
intended to be well-motivated physically, but rather to be simple and
flexible. The theoretical spectra modeled in \thomp, to which we
compare our results in \S\ref{s:disc}, are derived numerically and
hence do not suggest a more appropriate analytic expression for us to
employ. While \citet{c92} and others typically use broken power laws
to model aged synchrotron spectra, one of the key contentions of
\thomp\ is that other loss processes will significantly modify raw
synchrotron spectra in starbursts. We note that functional forms
analogous to ours have proven to be good empirical models in works
such as \citet{baars77} and the VLA broadband calibrator models. The
parabolic model has the additional benefits of being linear in its
parameters (in log-space) and being reducible to a pure power-law
spectrum (ignoring FFA) simply by fixing $C = 0$.

An alternative model for the effect of FFA upon the nonthermal
emission is one in which the absorption screens but is not mixed with
the nonthermal emitting medium,
\begin{equation}
\label{e:screen}
S_\mathrm{nt}(\tnu) = \tilde S_\mathrm{nt}(\tnu) \cdot e^{-\tff}.
\end{equation}
We disfavor this model for our targets because they are starbursts,
and their diffuse synchrotron emission propagates through a compact
ISM riddled with \HII\ regions and supernova remnants
\citep[\eg][regarding M82]{mmwpb02,rvzga04}. In the analysis that
follows we will, however, occasionally compare results for the two
absorption models.

\subsection{Isolating the thermal emission}
\label{s:isotherm}

Excepting those with active nuclei, most galaxies have a nonthermal
spectrum much steeper than the aforementioned $\alpha = -0.1$ of
optically-thin thermal free-free emission. Because the nonthermal
emission dominates at cm wavelengths, good higher-frequency data are
needed to usefully constrain the thermal component. However, above
$\nu \simeq 100$ GHz, thermal emission from cold dust begins
contributing non-negligibly to the total flux. We incorporated flux
densities from the literature with $20 < \nu / \textrm{GHz} < 100$
into our fits for this reason.  Table~\ref{t:hfobs} summarizes the
parameters of some of the relevant observations found in the
literature. Those observations, and others not listed in
table~\ref{t:hfobs}, are shown in Figures~\ref{f:s82}-\ref{f:s220} for
comparison to our observations.

Comparing fluxes in this way requires care because interferometers
resolve out spatially extended emission. The largest angular scale
(LAS) to which an interferometer observation is sensitive is inversely
proportional to its shortest baseline as measured in wavelengths. In a
very real way, two different interferometer observations, or an
interferometer observation and a single-dish observation, are not
``looking at'' the same source. We deal with these considerations on a
source-by-source basis below.

\subsubsection{M82}
\label{s:sf82}

M82 is well-observed in the 20--100 GHz range with both
interferometers and single dishes. The interferometric fluxes are
systematically smaller than the single-dish fluxes by about 25\%, with
the exception of the unusually low 87.2 GHz single-dish measurement of
\citet{jhm78}. Unlike the other single-dish data, which come from
maps, this observation was made with a single pointing of an 11 m dish
(HPBW 75\arcsec). It furthermore needs conversion to the flux density
scale of \citet{baars77}. We regard it as an outlier and discard
it.

All of the interferometer observations we consider
\citep{ck91,rvzga04,ck91,scbb96} have LASs of $\sim$50\arcsec. The
deconvolved size of the radio emission of M82 is $\sim$35\arcsec
 $\times$ $\sim$10\arcsec\ from cm wavelengths (this work) up to 92 GHz
\citep{kwm88,ck91}, so significant flux from M82 should not be
resolved out; the source of the excess emission detected by the
single-dish observations is unclear. For safety, we compare the ATA
fluxes only to other interferometric measurements. This criterion is
coincidentally equivalent to the selection of only post-1990
observations.

\subsubsection{NGC~253}
\label{s:sf253}

The extended spatial structure of the emission of NGC~253, as detected
by the ATA, makes the modeling of its spectrum more difficult. We do
not expect the high-frequency interferometer observations of NGC~253
listed in Table~\ref{t:hfobs} to be sensitive to the extended
component because of its steep spectral index and the constrained LASs
of the observations. Simple simulations of high-frequency VLA
observations of the galaxy support this conclusion. Two of the FIR
observations of NGC~253 \citep{rhla73,h+77} are made with single
dishes with $\sim$65\arcsec\ beams that could be sensitive to the
extended emission region; however, \citet{weiss08} resolve the central
component from the extended source and find that emission from the
former dominates the latter. Because of these facts, our spectra link
the FIR and high-frequency measurements with fluxes for only the
central component of NGC~253.

\subsubsection{Arp~220}
\label{s:sf220}

The observations of Arp~220 made by \citet{intok07}, \citet{n88} and
\citet{zagv96} have LASs that are too small for our purposes and are
not used in our fits. The 97 GHz single-dish observation of
\citet{avmgz00} is significantly discrepant from 87 GHz observations
made with the Plateau de Bure Interferometer \citep{r+91}, the
Nobeyama Millimeter Array \citep{intok07}, and a 110 GHz measurement
made with the Owens Valley Radio Observatory (OVRO) mm Array
\citep{ssss91}. The 113 GHz measurement also presented in
\citet{avmgz00} clusters much better with these other observations and
a power-law extrapolation of FIR observations of Arp~220
\citep{cksn92,de01,ewd89,rlr96,w+89}. Once again, we use the
interferometric observation but not the single-dish one.

\subsection{Modeling results}
\label{s:modres}

We performed least-squares fits between the averaged ATA data and the
chosen high-frequency measurements for each source, using a
Levenberg-Marquardt algorithm with $\tff$ constrained to be
nonnegative. In the case of M82 we used $\Omega = 2400$ arcsec$^2$,
its modeled angular size at 1 GHz. For the core component of NGC~253,
we used its modeled size as resolved at higher frequencies, 160
arcsec$^2$. The ATA did not resolve Arp~220 but in \citet{ssss91} it
is reported that the emission comes from an extended component of
$\sim$7\arcsec\ $\times\sim$15\arcsec\ and a stronger central
component of $\sim$2 arcsec$^2$. We adopt an intermediate value of 17
arcsec$^2$. In all cases we used $T_e = 10^4$ K. The numerical results
of our fits are reported in Table~\ref{t:fitdata}. The uncertainties
shown are marginalized values determined via Monte Carlo simulations
with 10000 realizations for each source. To provide a more physical
interpretation of $B$ and $C$, we also transform them, respectively, into $\alpha_5$,
the spectral index of the nonthermal model component at 5 GHz, and
$\Delta \equiv \textrm{d}\alpha/\textrm{d}(\log_{10} \nu)$ at the same
frequency. We also compute $f_\mathrm{th}$, the thermal fraction (not
correcting for FFA) at 1 GHz. We computed 3$\sigma$ confidence limits
in the total flux of the best-fit models at 40 frequencies
logarithmically spaced between 0.7 and 100 GHz. Finally, we computed
1$\sigma$ confidence regions in the $B$-$C$ plane of parameter space.
This was done by tracing the contour of $\chi^2 = \chi^2_\mathrm{min}
+ 2.3$, with $\chi^2_\mathrm{min}$ being the minimum $\chi^2$ achieved
in the overall fit and the offset parameter coming from \citet{a76}.
The value of $\chi^2$ assigned to each location in the $B$-$C$ plane
is the minimum value of $\chi^2$ attained in a fit with $B$ and $C$
fixed and $A$ and $\tau_1$ free.

The parameter values that are obtained depend strongly on the
highest-frequency data points. Fits that were run without any very
high-frequency ($\nu > 50$ GHz) data points often converged to the
physically-suspect lower bound of $\tau_1 = 0$. To investigate the
possible effects of inaccurate high-frequency observations on our
results, we ran a set of Monte Carlo simulations in which the
uncertainty associated with each high-frequency measurement was
doubled. In this case, the resulting uncertainties in the modeled
parameter values increased by a factor of $\sim$1.5, demonstrating the
extent to which the results are affected by those few data points.
While we have done our best to choose our high-frequency data points
carefully, it remains true that our results are dependent on the
choice of which measurements to include and, of course, the accuracy
of those measurements.

\subsubsection{M82}
\label{s:mr82}

The M82 fit has $\srcs = 1.21$. The ATA data and fit results are
presented in Figure~\ref{f:s82}. The value of $\alpha_5$ that we
obtain agrees well with that found in \citet{kwm88}, $\alpha = -0.68
\pm 0.02$. We, however, find a higher thermal content at 32 GHz,
because the possibility of a steepening nonthermal spectrum allows for
a larger thermal contribution.

A fit with a fixed value of $C = 0$ (\ie, a pure power-law nonthermal
spectrum) had a comparable $\srcs = 1.34$, but also $\tau_1 = 0$, a
physically implausible result. In the $C = 0$ case, the many ATA data
points tightly constrain the power-law slope of the nonthermal
component, but the high-frequency points lie significantly below the
nonthermal line. The thermal component, being spectrally shallower
than the nonthermal component, can only make this situation worse, so
the fit converges to a solution with no thermal component at all. If
any thermal component is present, the nonthermal spectrum must steepen
so that the high-frequency points can be modeled successfully.

\subsubsection{NGC~253}
\label{s:mr253}

\begin{figure}[p]
\plotone{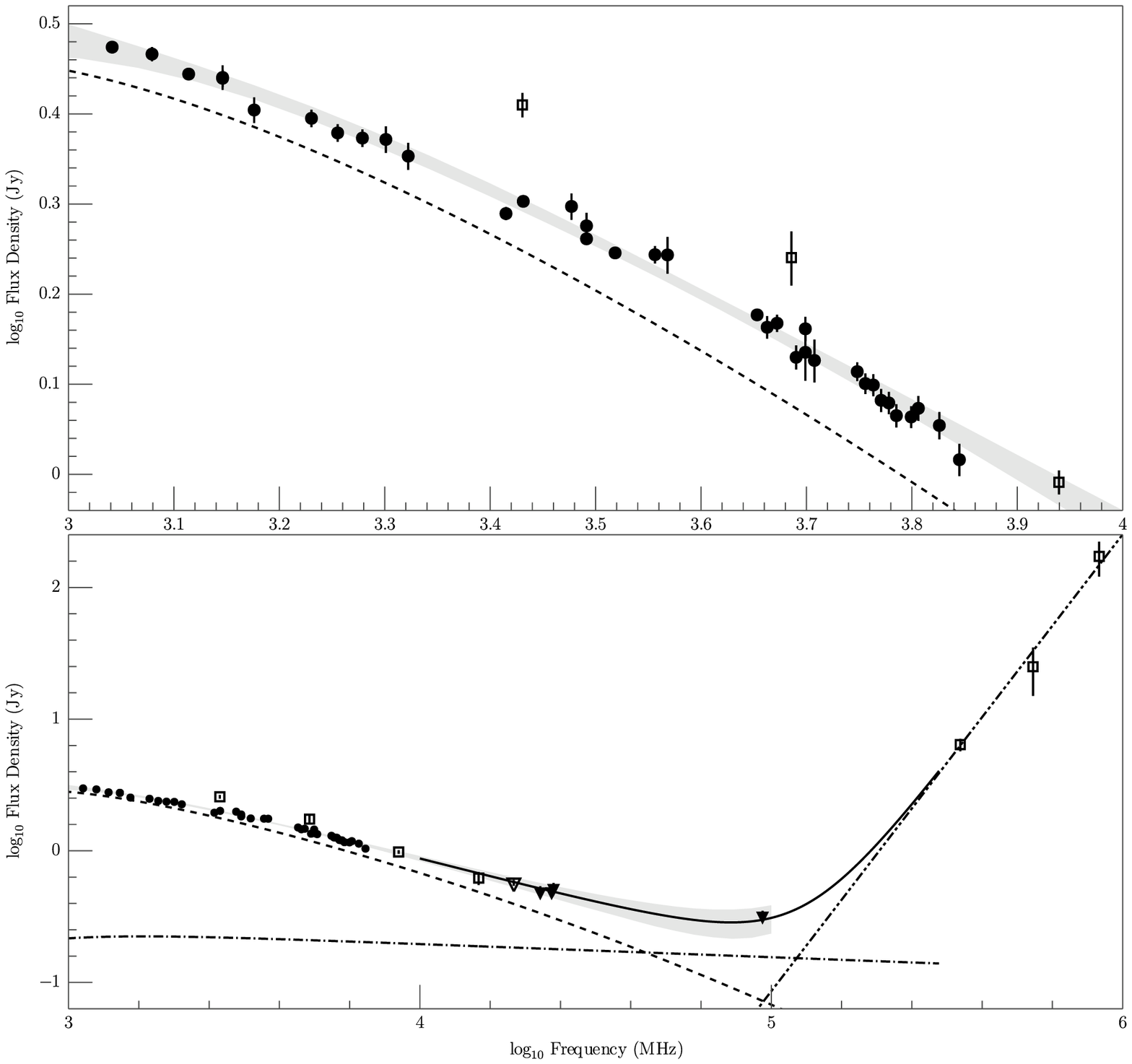}
\caption{Spectrum of the core component of NGC~253 with model. The
  symbols are as in Figure~\ref{f:s82}. External data are found in
  \citet{bbew79,h+77,owhw05,pzwls96,rpgsz06,rhla73,thwnk05,weiss08}.
  See \S\ref{s:mr253} for discussion.}
\label{f:s253_1}
\end{figure}

\begin{figure}[p]
\plotone{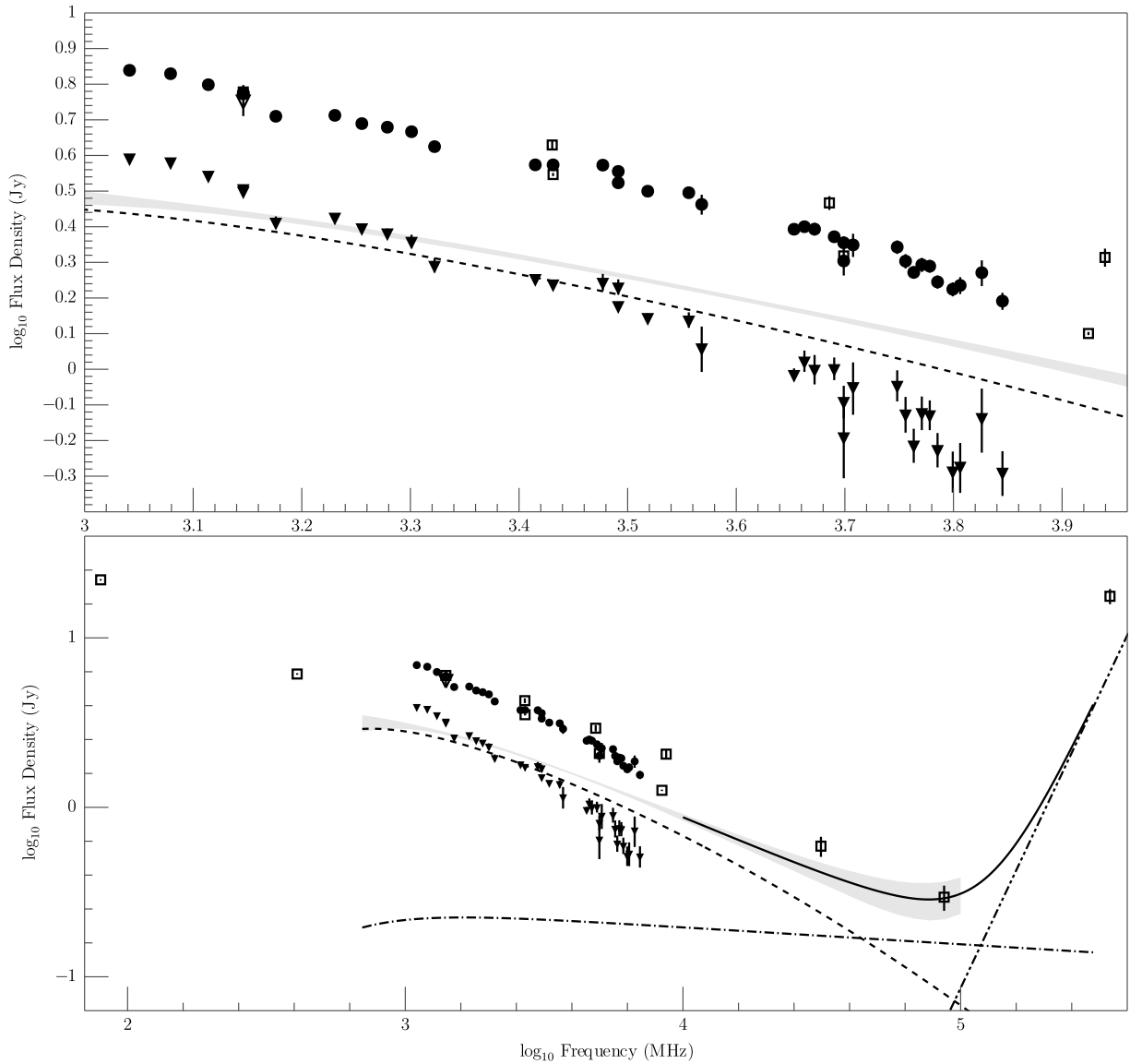}
\caption{Spectra of the extended component and total flux of NGC~253.
  The model used in Figure~\ref{f:s253_1} is overlaid to ease
  comparison. External data are found in
  \citet{bbew79,gw81,jhm78,parkes,shchw04,weiss08}. {\it Triangles:}
  fluxes of the extended component of NGC~253. {\it Circles:} sum of
  extended and core component fluxes. Other symbols are as in the
  previous plot.}
\label{f:s253_500}
\end{figure}

\begin{figure}[p]
\plotone{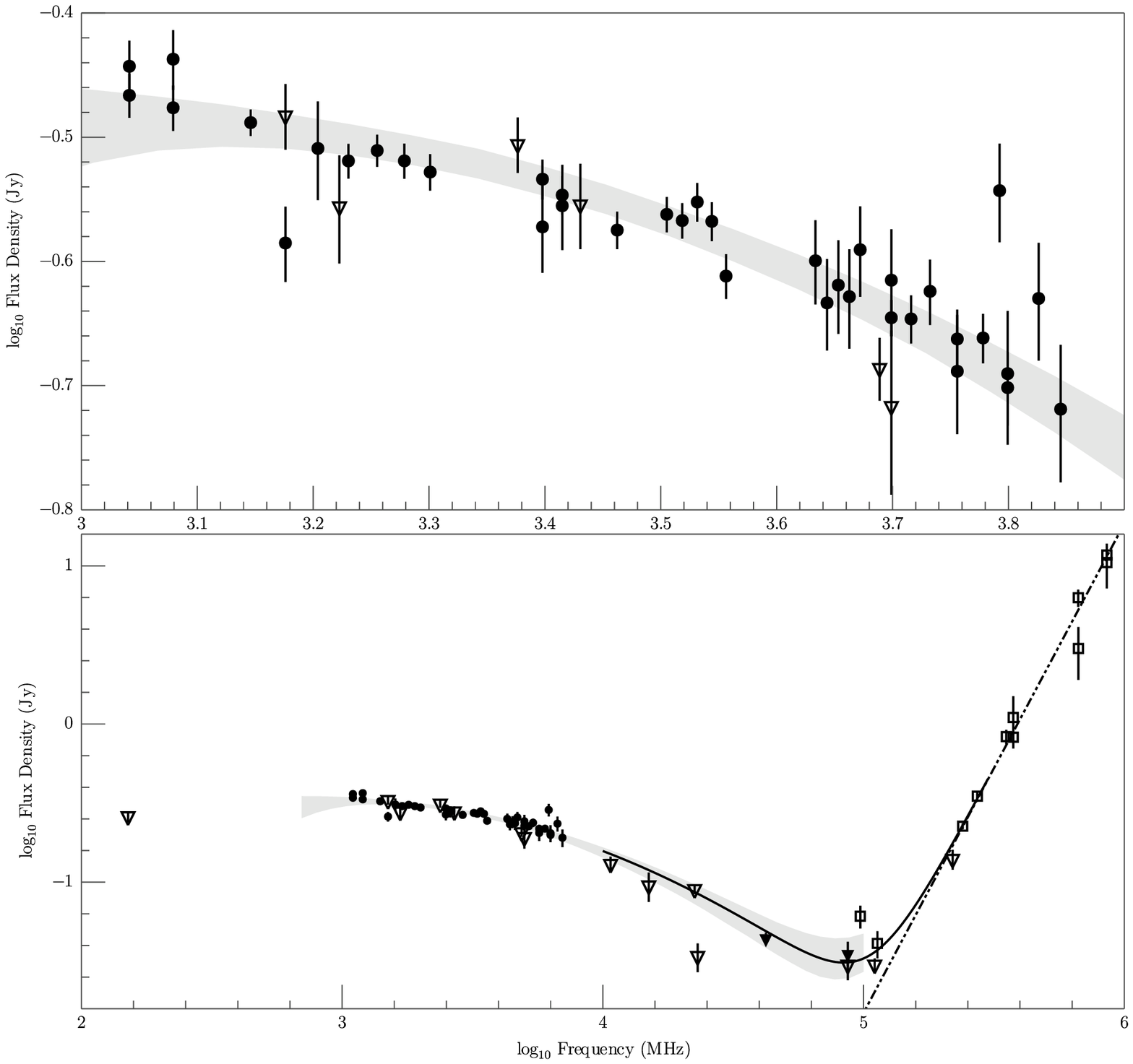}
\caption{Spectrum of Arp~220 with model. The symbols are as in
  Figure~\ref{f:s82}. See \S\ref{s:mr220} for further discussion.
  External data are found in
  \citet{avmgz00,cksn92,de01,ewd89,intok07,n88,r+91,rlr96,r+05,ssss91,sa91,w+89,zagv96}.}
\label{f:s220}
\end{figure}

\begin{figure}[p]
\centering
\includegraphics[width=4in]{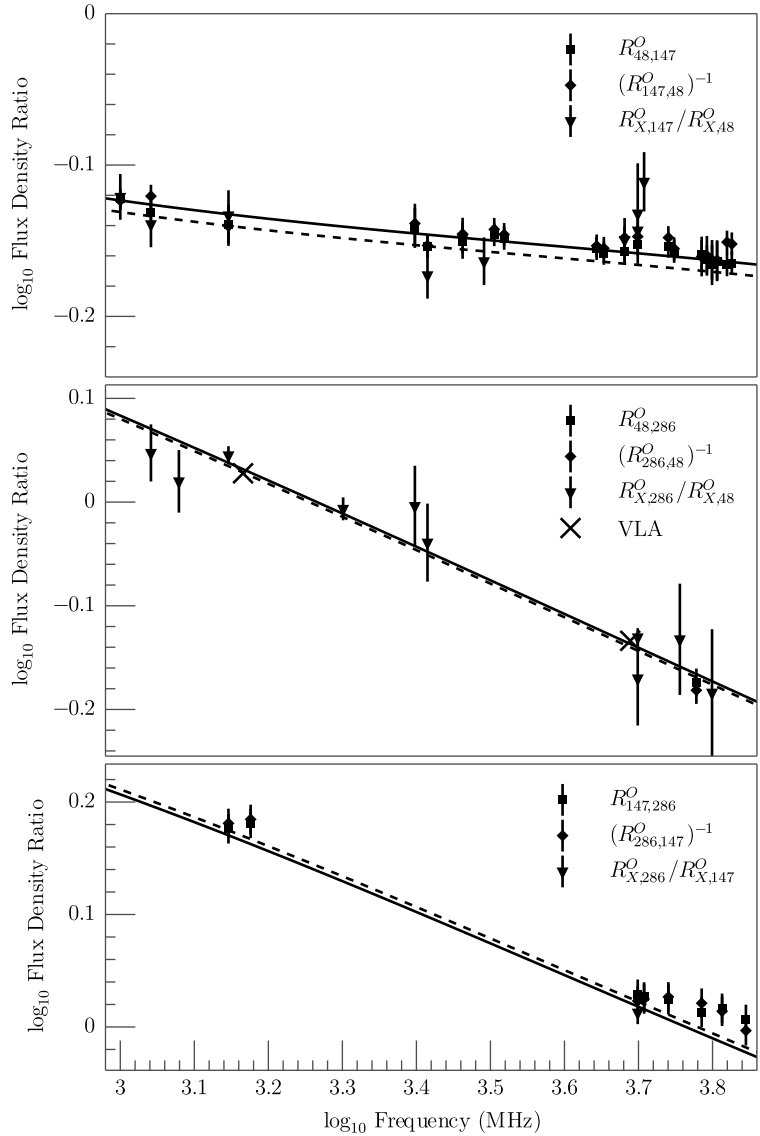}
\caption{Flux density ratios of calibrator observations and spectral
  models. See \S\ref{s:modcheck} for definitions. {\it Dashed lines:}
  ratios of the VLA 1999.2 models. {\it Solid lines:} ratios of the
  spectral models adopted in this work. {\it Squares and diamonds:}
  directly observed calibrator ratios. The good agreement between the
  original and inverted ratios indicates that there is negligible flux
  loss in the secondary calibration process (\S\ref{s:copycal}). {\it
    Triangles:} calibrator ratios computed indirectly from the ratio
  of two observations of a science target. {\it Xs:} measurements of
  $R^O_{48,286}$ made at the VLA in 2008 September. The observations
  of $R^O_{147,286}$ were made in 2009 June and exhibit discrepant
  behavior; see \S\ref{s:modcheck} for discussion.}
\label{f:consistency}
\end{figure}

The NGC~253 fit has $\srcs = 1.02$. The ATA data and model results for
the core component of NGC~253 are presented in
Figure~\ref{f:s253_1}. The symbols used are analogous to those in
Figure~\ref{f:s82}. Figure~\ref{f:s253_500} shows the extended and
total fluxes in context with total-flux data from the literature. The
same models as in Figure~\ref{f:s253_1} are overlaid to ease
comparison.

A fit with a fixed value of $C = 0$ had $\srcs = 1.51$ and $\tau_1 =
0.25$. Here, a change in the spectral index is detectable in the ATA
data alone, so when the nonthermal component has a constant slope,
$\tau_1$ must increase to induce more curvature at low frequencies via
FFA. In this case, once again the high-frequency data points lie well
below the points attained by the model.

\subsubsection{Arp~220}
\label{s:mr220}

The Arp~220 fit has $\srcs = 1.23$ and $\tau_1 = 0$ and is shown in
Figure~\ref{f:s220}. This result is due to the dramatic steepening of
the spectrum of Arp~220 at $\nu \sim 10$~GHz. A purely nonthermal fit
that accommodates the high-frequency observations must underestimate
the low-frequency ($\nu \lesssim 2$~GHz) data points significantly.
The addition of any thermal contribution would require an even steeper
nonthermal component to match the high-frequency measurements, which
would make the discrepancies on the low end even worse. Adjustments to
the assumed angular size of Arp~220, which affect the overall
scaling of the thermal component, do not improve the outcome. We
report the results of our fit in Table~\ref{t:fitdata} but caution
that they are unsatisfactory. We exclude the Arp~220 results from
further analysis in the following sections.

If the nonthermal curvature parameter $C$ is fixed to zero, the model
cannot accommodate the data, with $\srcs = 2.77$. This result
quantitatively confirms the supposition of \citet{c92} that the
nonthermal spectrum of Arp~220 steepens in the cm regime. Our data
cause serious problems for the empirical spectral model of Arp~220
presented in \citet{sa91}, which predicts a much steeper spectrum at
frequencies of a few GHz than we find. Their model was inconsistent
with several measurements at $\nu \sim 2$~GHz shown in the paper, and
our ATA data make it clear that those measurements are generally
accurate and the simple power-law used in that work is insufficient to
match the data.

\section{Discussion}
\label{s:disc}

We test whether our observational results for M82 and NGC~253 are
consistent with the predictions of \thomp\ by comparing the curvature
of the FFA-corrected nonthermal spectra with the output of the
numerical model presented in that work. We obtained an updated version
of the relevant code (E. Quataert, private communication) and modified
it to allow more convenient tuning of the input parameters. These
parameters are (1) $\Sigma_g$, the surface density of the gas disk;
(2) \heff, the effective scale height of the gas disk; (3) $p$, the
power-law index of the injected CR energy spectrum, $\mathrm{d}n
\propto E^{-p} \mathrm{d}E$; and (4) $\eta$, the ratio of the energy
density of the magnetic field to the pressure of the ISM, $B^2 = \eta
\cdot 8 \pi^2 G \Sigma_g^2$. The value of \heff\ affects the
characteristic timescales of CR cooling via ionization and free-free
losses by setting the density of the neutral ISM as encountered by the
CRs, $n_\mathrm{CR} = \Sigma_g / 2 m_p \heff$. (In \thomp, \heff\ is
specified by two parameters, the physical scale height of the gas disk
$h$ and an ISM/CR interaction efficiency parameter $f$: $\heff = h /
f$. They are degenerate for our purposes.) Because the mean value of
the ISM density may not correspond well to the value of
$n_\mathrm{CR}$ due to clumping and other effects, \heff\ may likewise
not correspond well to the actual scale height of the gas disk. We
take the surface density of the disk of M82 to be 0.081~g~cm$^{-2}$
and that of NGC~253 to be 0.067~g~cm$^{-2}$ \citep[and references
therein]{t09}. Note that these recently-derived values are
significantly different than the ones given in \citet{k98}.

Given a choice of $\Sigma_g$, \heff, $p$, and $\eta$, we used a simple
least-squares fit to determine values for the spectral parameters $B$
and $C$. We used the \thomp\ model to compute flux densities
at 25 frequencies logarithmically spaced between 1 and 7 GHz and then
fit a bent power law (as in Equation~\ref{e:nonth}) to the data. All
of the spectra generated in this way were very well-described by such
a model. The derived parameters were not sensitive to the details of
the choice of frequencies at which to sample the model.

\begin{figure}[p]
\plotone{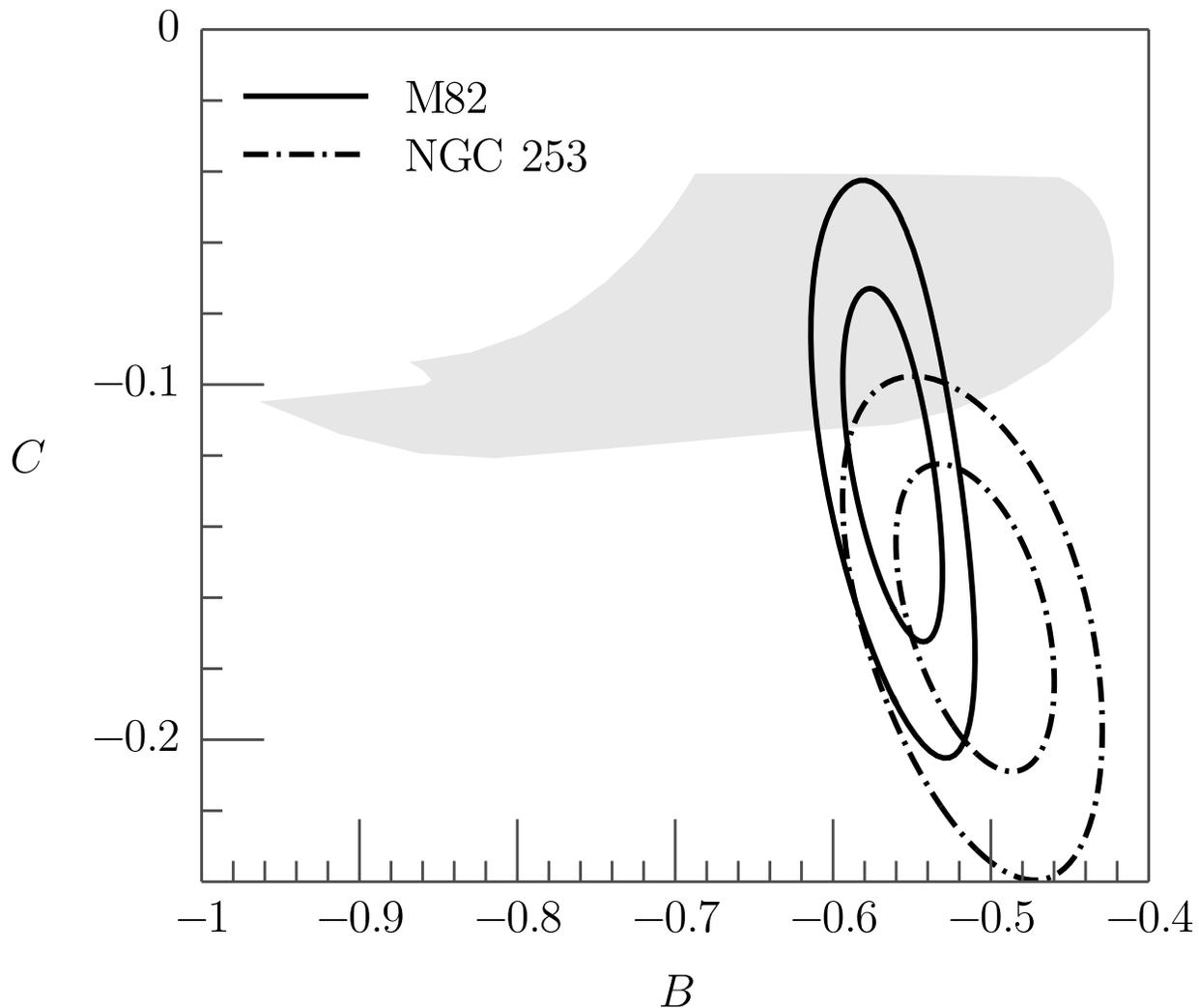}
\caption{Observationally and theoretically derived values of the
  spectral curvature parameters $B$ and $C$. {\it Contours:} 1$\sigma$
  ({\it inner}) and 2$\sigma$ ({\it outer}) confidence regions for the
  parameters as derived from the modeling described in
  \S\ref{s:model}. {\it Shaded region:} the range of $B$ and $C$
  values taken by the model of \thomp\ over a range of parameters
  appropriate for M82 as described in \S\ref{s:disc}. The equivalent
  region for the NGC~253 parameters is virtually identical to the one
  shown.}
\label{f:bcvals}
\end{figure}

We computed such $B$ and $C$ values for the \thomp\ model on two grids
of parameters. The first had $\Sigma_g$ set to the observed M82 value,
\heff\ ranging between 10 and 250 pc, $p$ ranging between 2 and 2.5,
and $\eta$ ranging between 0.02 and 1, all stepping through 10 values
logarithmically. The lower bound of $\eta$ is the approximate value
for which the synchrotron cooling time, $\tau_\mathrm{syn}$, and the
timescale for escape from the galaxy via advection by a galactic wind,
$\tau_\mathrm{esc}$, are equal. For $\eta$ (and hence $B$) smaller
than this, the assumptions of the \thomp\ code are no longer upheld.
The largest plausible $\eta$ is $\sim$1, at which point the magnetic
field and ISM pressure are in equipartition \citep{p66}. The second
grid was identical except that $\Sigma_g$ was set to the NGC~253 value
and the lower bound of $\eta$ was 0.04, the analogous limiting value.
Figure~\ref{f:bcvals} shows the values of $B$ and $C$ attained by the
model and compares them to the 1$\sigma$ and 2$\sigma$ confidence
intervals associated with our observations. The \thomp\ models can
accommodate a wide range of spectral indices and exhibit spectral
steepening for all parameter values. Our observed spectra, however,
exhibit more steepening (more negative $C$) than the models. The
models are consistent with observations of both galaxies at the
2$\sigma$ level but only with M82 at the 1$\sigma$ level. If the
source of the FFA is configured as a screen in front of the nonthermal
emission rather than a medium mixed within it
(Equation~\ref{e:screen}), the $C$ values derived for our observations
are less negative, $\sim-0.10$, and the \thomp\ models have large
areas of overlap with the observations of both galaxies even at the
1$\sigma$ level. The \srcs\ values of the fits in this case are
virtually identical to those of the fits in the well-mixed case.

\begin{figure}[p]
\plotone{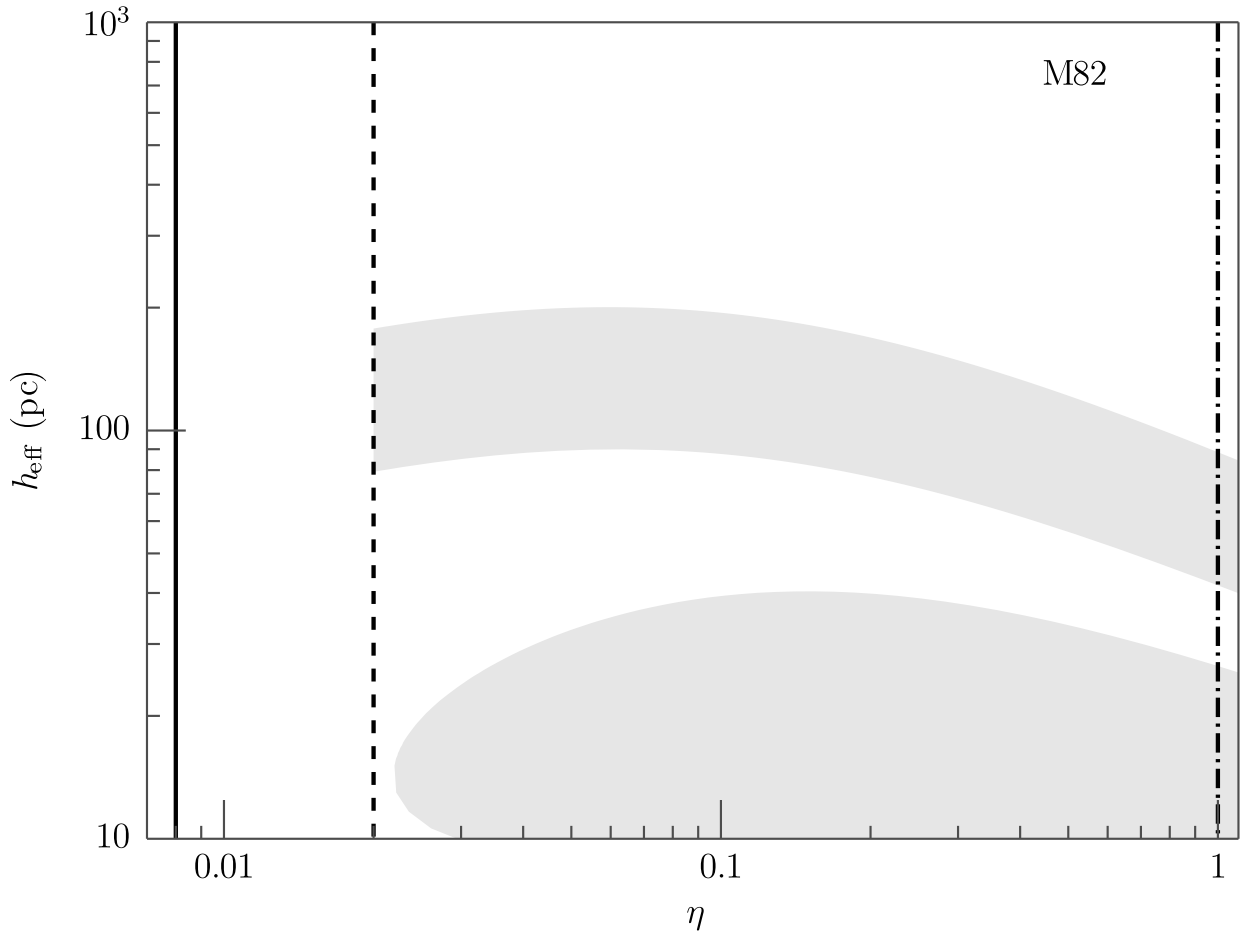}
\caption{Values of $\eta$ and \heff\ for which spectral curvature
  parameters $B$ and $C$ derived from the \thomp\ model are consistent
  with observations of M82 to within 2$\sigma$. {\it Upper shaded
    region:} the allowed parameters for $p = 2$. {\it Lower shaded
    region:} the allowed parameters for $p = 2.3$. {\it Solid line:}
  the value of $\eta$ such that $B = B_\mathrm{min}$. {\it Dashed
    line:} the value of $\eta$ such that the synchrotron cooling
  timescale $\tau_\mathrm{syn}$ and the escape time
  $\tau_\mathrm{esc}$ of \thomp\ are equal, assuming $\nu = 1.43$~GHz,
  $\heff = 100$~pc, and the galactic wind speed $v_w =
  500$~km~s$^{-1}$. This defines the approximate lower limit in $\eta$
  of the validity of the \thomp\ numerical model. {\it Dot-dashed
    line:} the value of $\eta$ such that $B = B_\mathrm{eq}$.}
\label{f:hetac82}
\end{figure}

\begin{figure}[p]
\plotone{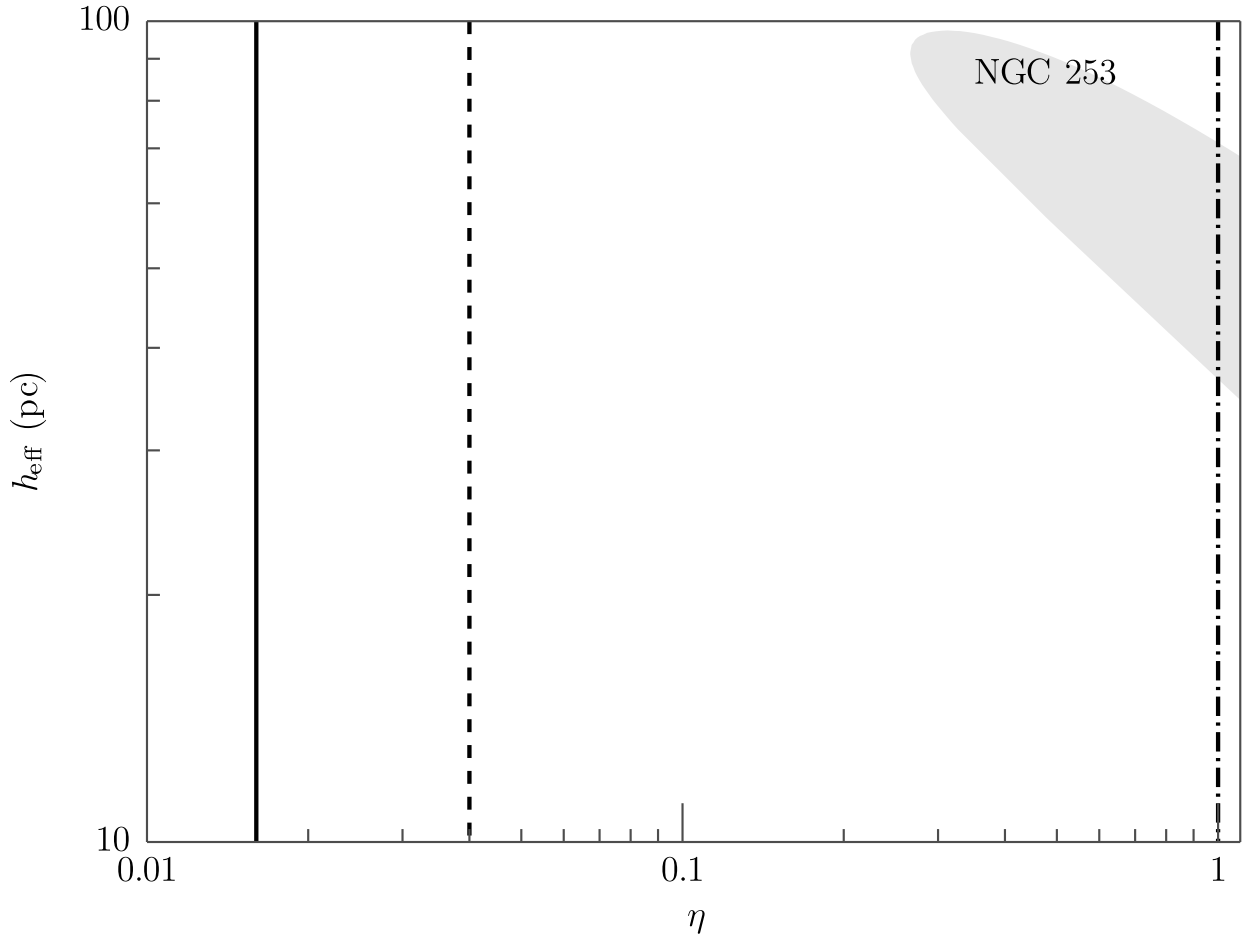}
\caption{Values of $\eta$ and \heff\ for which spectral curvature
  parameters $B$ and $C$ derived from the \thomp\ model are consistent
  with observations of NGC~253 to within 2$\sigma$. The symbols are as
  in the previous figure. {\it Shaded region:} the allowed parameters
  for $p = 2$. The \thomp\ model with $p = 2.3$ is excluded to a
  99.99\% confidence limit.}
\label{f:hetac253}
\end{figure}

We quantify this agreement in Figures~\ref{f:hetac82} and
\ref{f:hetac253}, which show the regions in the $\eta$-\heff\ plane
where the results of the \thomp\ model are consistent with the
observed properties of M82 and NGC~253, respectively, at the 2$\sigma$
level, for $p = 2$ and $2.3$. Because our data, unlike the outputs of
the \thomp\ models, are affected by FFA and thermal emission, they
cannot be compared to the models directly without taking into account
these effects. To do so requires some sort of spectral modeling, so we
compare the spectral curvature parameters $B$ and $C$ found for our
data with ones derived for the outputs of the \thomp\ models. (If we
corrected our observations for the modeled FFA and thermal emission
and then compared them to the \thomp\ models, we would be effectively
hiding the model-dependence of the correction and discarding the
information that the data have been corrected assuming an underlying
parabolic nonthermal spectral model.) We assumed a bivariate normal
probability distribution for the observed values of $B$ and $C$, with
the parameters of the distribution estimated by applying the maximum
likelihood method to the results of the Monte Carlo simulations. The
agreement between the shape of the analytic distribution function and
contours of constant $\Delta\chi^2$ was very good.
Using the bounds on \heff\ and $\eta$ given above, $p = 2.5$ is
rejected at the 99.96\% confidence limit for M82 and the 99.99\%
confidence limit for NGC~253; larger values of $p$ become acceptable
if the lower limit on \heff\ is decreased. For NGC~253, $p = 2.3$ is
also rejected at the 99.99\% confidence limit.

For M82, virtually all physically plausible values of $\eta$ are
allowed. For $p = 2$, the allowed values of \heff\ are of order
100~pc, broadly consistent with observed radio sizes. For $p = 2.3$,
agreement requires $\heff \approx 20$~pc.  Such sizes have been
proposed for the neutral gas disks of some ultraluminous infrared
galaxies \citep[ULIRGs;][]{syb97,cwu98}, but are unlikely to be found
in less extreme galaxies. Furthermore, if the CRs interact with the
ISM at somewhat less than its nominal mean density (the parameter $f$
of \thomp\ is $<$1), as is suggested by measurements of CRs in the
Milky Way \citep{c98}, the actual scale height of the disk becomes
even less physically plausible. As $\eta$ increases, the allowed
\heff\ decreases for both values of $p$. This is because the stronger
magnetic field steepens the synchrotron spectrum, requiring a smaller
\heff\ to increase $n_\mathrm{CR}$ and consequently the amount of
spectral flattening due to ionization and free-free emission off of
neutral hydrogen. As $p$ increases, the allowed values of \heff\
rapidly decrease for the same reason.

The NGC~253 data are only marginally consistent with the \thomp\ model
within the range of parameters we consider, and the statistically
acceptable results have $\heff \approx 50$~pc. If the source of FFA is
configured as a screen rather than mixed with the nonthermal emission,
much more of the \thomp\ model parameter space is allowed, with a
general configuration of allowed parameters similar to that obtained
for M82.

The relationship between \heff\ and the observed radio sizes of
starburst galaxies $h_\mathrm{obs}$ is significant within the \thomp\
model.  Estimates in \thomp\ indicate that if $\heff \sim
h_\mathrm{obs}$ and $B \sim B_\mathrm{min}$, free-free emission off of
neutral ISM nuclei becomes a significant sink of CR energy for
$\Sigma_g \gtrsim 1$ g cm$^{-2}$, which would disrupt the observed
linearity of the FIR-radio correlation at high surface densities.
Since the free-free timescale is independent of the magnetic field
while the synchrotron timescale goes as $\tau_\mathrm{syn} \propto
B^{-3/2}$, a stronger magnetic field, $B \sim B_\mathrm{eq}$, avoids
this inconsistency.  Our observations have $\heff \sim h_\mathrm{obs}$
in the $p = 2$ case and $\heff \ll h_\mathrm{obs}$ in the $p = 2.3$
case. A precise independent measurement of $p$ could thus provide a
valuable test of this aspect of the \thomp\ model.

\section{Conclusions}
\label{s:conc}

Our observations of M82 are consistent with the scenario proposed in
\thomp. Our observations of NGC~253, on the other hand, are only
marginally consistent with it. We decisively detect the spectral
steepening at $\nu \simeq 1$ GHz predicted in that work. The
uncertainties in our spectral modeling, due to uncertainties in the
individual observations, the difficulties of untangling the thermal
and nonthermal emission, and the inaccuracies inherent in our
spatially-unresolved approach, make it difficult to significantly
constrain all of the parameters of the \thomp\ spectral model or to
reject its entire parameter space. We do, however, reject with high
confidence the \thomp\ model with a steep electron energy injection
spectrum power-law index of $p = 2.5$ for M82 and NGC~253. We also
reject with high confidence the \thomp\ model with $p = 2.3$ for
NGC~253. We are unable to place limits on the strength of the magnetic
field required by the \thomp\ model with our current data. Further
work to constrain the efficiency with which CRs interact with the ISM
or the allowed values of $p$ would significantly shrink the model
parameter space. Observations of a larger sample of galaxies would
allow a statistical treatment of the topics that we have dealt with
here on a case-by-case basis. It would likely also be fruitful to
compare our detailed radio results with FIR observations. Such an
approach would be a frequency-space analog to the recent works
investigating the connections between FIR and radio emission at high
spatial resolutions \citep{mh95,p+06}.

The broadband, continuous spectra that we have obtained with the ATA
are precise and repeatable. The quality of our measurements is limited
by several factors. The \uv\ coverage of our snapshot observations is
good but suboptimal for complex sources, which made the imaging of
targets such as NGC~253 more difficult. The sensitivity the system is
a limiting factor for snapshot observations of fainter sources such as
Arp~220. Both of these quantities were limited by the limited number
of antennas available for observing during the commissioning of the
ATA-42. The planned expansion of the ATA to $\sim$350 antennas will do
much to improve both of them. The fact that the ATA was undergoing
commissioning at the time of these observations meant that several
important aspects of observing and data reduction, such as
polarization calibration and RFI mitigation, were incomplete.  Ongoing
development since these observations were made has led to improvements
in virtually all aspects of the system performance. Of particular
interest is the deployment of a second independently-tunable
correlator, which doubles the already-fast throughput of our spectral
observations and further improves the prospects for fast, precise,
broadband spectral monitoring of sources.

\acknowledgements

The authors would like to acknowledge the generous support of the Paul
G. Allen Family Foundation, who have provided major support for
design, construction, and operations of the ATA. Contributions from
Nathan Myhrvold, Xilinx Corporation, Sun Microsystems, and other
private donors have been instrumental in supporting the ATA. The ATA
has been supported by contributions from the US Naval Observatory in
addition to National Science Foundation grants AST-050690 and
AST-0838268.

Our work would not have been possible if not for the tireless efforts
of everyone on the ATA team: Jack Welch, Don Backer, Leo Blitz,
Douglas Bock, Calvin Cheng, Steve Croft, Matt Dexter, Greg Engargiola,
Ed Fields, R. James Forster, Colby Gutierrez-Kraybill, Carl Heiles,
Tamara Helfer, Susanne Jorgensen, Garrett Keating, Casey Law, John
Lugten, Dave MacMahon, Peter McMahon, Oren Milgrome, Andrew Siemion,
Douglas Thornton, Lynn Urry, Joeri van Leeuwen, Dan Werthimer, Melvyn
Wright; Jill Tarter, Rob Ackermann, Shannon Atkinson, Peter Backus,
Billy Barott, Tucker Bradford, Mike Davis, Dave DeBoer, John Dreher,
Gerry Harp, Jane Jordan, Tom Kilsdonk, Tom Pierson, Karen Randall,
John Ross, Seth Shostak, Ken Smolek; Matt Fleming, Chris Cork, Artyom
Vitouchkine; Niklas Wadefalk, and Sandy Weinreb.

We thank Eliot Quataert for helpful discussions. We are indebted to
the numerous programmers and scientists who have contributed to the
software systems used in this work, the entirety of which are
open-source and freely available for inspection and use by anyone in
the world. This research has made use of NASA's Astrophysics Data
System; the SIMBAD database, operated at CDS, Strasbourg, France; and
the NASA/IPAC Extragalactic Database (NED) which is operated by the
Jet Propulsion Laboratory, California Institute of Technology, under
contract with the National Aeronautics and Space Administration. PKGW
is supported by an NSF Graduate Research Fellowship.

\clearpage


%
%
%

\begin{deluxetable}{rc@{\extracolsep{\otofs}}p{1.8in}p{2in}}
\tablecaption{Observations\label{t:obs}}
\tablehead{
  \colhead{Date (UT)} &
  \colhead{Segment \#} &
  \colhead{Sources} &
  \colhead{Frequencies (GHz)}
}
\startdata
20080912 & 1 
& \otd 3C~48, 3C~147, M82, NGC~253
& \otd 1.4, 1.5, 2.0, 2.1, 3.1, \mbox{4.0\tablenotemark{a},}
\mbox{4.1\tablenotemark{a},} 5.0, 5.1 \\

20080913 & 2 
& \otd 3C~286, Arp~220
& \otd 1.5, 1.6, 2.5, 2.6, 4.5, 4.6 \\

& 3 
& \otd 3C~147, NGC~253
& \otd 1.1, 1.2, 1.3, 2.6, 2.7, 3.6, \mbox{3.7\tablenotemark{a},}
  5.6, 5.7 \\

 & 4 
& \otd 3C~48, 3C~147, M82
& \otd 3.2, 3.3, 4.4, 4.5, 5.5, 5.6, 6.6, 6.7 \\

 & 5 
& \otd 3C~286, Arp~220
& \otd 1.1, 1.2, 2.5, 2.6, 5.0, 5.7, 6.3 \\

20080914 & 6 
& \otd 3C~48, Arp~220
& \otd 1.1, 1.2, 2.5, 2.6, 5.0, 5.7, 6.3 \\

 & 7 
& \otd 3C~147, NGC~253
& \otd 2.7, 3.1, 3.3, 4.5, 4.7, 6.1, 6.3 \\

 & 8 
& \otd 3C~48, 3C~147, M82
& \otd 1.0, 1.1, 2.6, 2.9, 4.8, 5.0 \\

 & 9 
& \otd 3C~286, Arp~220
& \otd 1.9, 2.9, 3.2, 3.3, 5.4 \\

20080924 & 10 
& \otd 3C~286, Arp~220, M82
& \otd 1.4, 1.7, 1.8, 2.0, 3.4, 3.5, 3.6, 5.0, 5.2, 6.0 \\

20081015 & 11 
& \otd 3C~48, 3C~286, Arp~220
& \otd 0.5, 0.7, 0.9, 1.4, 4.3, 4.4, 4.7, 5.0, 6.2, 6.7, 7.0 \\

 & 12 
& \otd 3C~48, 3C~147, M82
& \otd 1.4, 2.4, 2.5, 4.3, 4.7, 5.0 \\

 & 13 
& \otd 3C~147, NGC~253
& \otd 1.4, 1.7, 1.8, 1.9, 4.5, 4.6, 4.9, 5.0, 5.8, 5.9,
  6.0, 6.4, 6.7, 7.0 \\

 & 14 
& \otd 3C~48, 3C~147, M82
& \otd 6.1, 6.2, 6.3, 6.4 \\

20090307 & 15 
& \otd 3C~48, 3C~286
& \otd 1.4, 3.1, 5.0, 6.0, 7.0 \\

20090621 & 16 
& \otd 3C~147, 3C~286
& \otd 1.0, 1.4, 1.5, 5.0, 5.1, 5.5, 6.1, 6.5, 7.0

\enddata
\tablenotetext{a}{Discarded due to severe RFI.}
\end{deluxetable}

\begin{deluxetable}{lccccc}
\tablecaption{Snapshots made during segment \#12\label{t:oscans}}
\tablehead{
  \colhead{Time} &
  \colhead{Pointing\tablenotemark{a}} &
  \colhead{Azimuth} &
  \colhead{Elevation} &
  \colhead{$\nu_\mathrm{sky}$} &
  \colhead{Focus\tablenotemark{a}} \\
  (UT) & & (deg) & (deg) & (GHz) & (GHz)
}
\startdata
03:50 & 3C~48  & 77 & 39 & 1.4 & 2.0 \\
03:53 &        & 77 & 40 & 2.4 & 3.0 \\
03:55 &        & 78 & 40 & 2.5 &     \\
03:59 & M82    & 359 & 21 & 1.4 & 2.0 \\
04:02 &        & 358 & 21 & 2.4 & 3.0 \\
04:04 &        & 358 & 21 & 2.5 &     \\
04:08 & 3C~48\tablenotemark{b} & 79 & 43 & 4.3 & 5.0 \\
04:11 &        & 80 & 43 & 4.7 &     \\
04:13 &        & 80 & 44 & 5.0 &     \\
04:17 & M82    & 359 & 21 & 4.3 &     \\
04:20 &        & 0 & 21 & 4.7 &     \\
04:22 &        & 0 & 20 & 5.0 &     \\
04:28 & 3C~48\tablenotemark{b}  & 82 & 47 & 1.4 & 2.0 \\
04:31 &        & 82 & 47 & 2.4 & 3.0 \\
04:34 &        & 83 & 48 & 2.5 &     \\
04:38 & M82    & 1 & 21 & 1.4 & 2.0 \\
04:41 &        & 2 & 21 & 2.4 & 3.0 \\
04:43 &        & 2 & 21 & 2.5 &     \\
04:47 & 3C~147 & 40 & 19 & 1.4 & 2.0 \\
04:50 &        & 40 & 19 & 2.4 & 3.0 \\
04:52 &        & 40 & 20 & 2.5 &     \\
04:56 & 3C~48  & 86 & 52 & 4.3 & 5.0 \\
04:58 &        & 86 & 52 & 4.7 &     \\
05:01 &        & 87 & 53 & 5.0 &     \\
05:04 & M82    & 4 & 21 & 4.3 &     \\
05:07 &        & 4 & 21 & 4.7 &     \\
05:09 &        & 4 & 21 & 5.0 &     \\
\enddata
\tablenotetext{a}{A blank entry indicates no change from the previous
  value.}
\tablenotetext{b}{Had it been above the 18$^\circ$ elevation limit of the 
  ATA, 3C~147 would have been observed at these times.}
\end{deluxetable}


\begin{deluxetable}{ccc}
\tablecaption{Flux density measurements of M82\label{t:m82}}
\tablehead{
  \colhead{$\nu$ (GHz)} & \colhead{Flux (Jy)} & \colhead{Calibrator}
}
\startdata
$1.0$ & $8.92 \pm 0.27$ & 3C 48 \\
 & $8.96 \pm 0.16$ & 3C 147 \\
$1.1$ & $8.72 \pm 0.26$ & 3C 48 \\
 & $8.46 \pm 0.15$ & 3C 147 \\
$1.4$ & $7.25 \pm 0.11$ & 3C 48 \\
 & $7.37 \pm 0.10$ & 3C 286 \\
$1.5$ & $6.85 \pm 0.15$ & 3C 48 \\
$1.7$ & $6.60 \pm 0.09$ & 3C 286 \\
$1.8$ & $6.34 \pm 0.09$ & 3C 286 \\
$2.0$ & $6.06 \pm 0.13$ & 3C 48 \\
 & $6.13 \pm 0.08$ & 3C 286 \\
$2.1$ & $5.87 \pm 0.13$ & 3C 48 \\
$2.5$ & $5.39 \pm 0.12$ & 3C 48 \\
$2.6$ & $5.55 \pm 0.17$ & 3C 48 \\
 & $5.21 \pm 0.09$ & 3C 147 \\
$2.9$ & $4.89 \pm 0.11$ & 3C 48 \\
 & $4.88 \pm 0.11$ & 3C 147 \\
$3.0$ & $4.33 \pm 0.09$ & 3C 48 \\
$3.1$ & $4.70 \pm 0.10$ & 3C 48 \\
$3.2$ & $4.54 \pm 0.07$ & 3C 147 \\
$3.3$ & $4.50 \pm 0.07$ & 3C 147 \\
$3.4$ & $4.44 \pm 0.06$ & 3C 286 \\
$3.5$ & $4.32 \pm 0.06$ & 3C 286 \\
$3.6$ & $4.15 \pm 0.06$ & 3C 286 \\
$4.3$ & $3.85 \pm 0.09$ & 3C 48 \\
$4.4$ & $3.73 \pm 0.06$ & 3C 147 \\
$4.5$ & $3.67 \pm 0.06$ & 3C 147 \\
$4.7$ & $3.55 \pm 0.08$ & 3C 48 \\
$4.8$ & $3.53 \pm 0.08$ & 3C 48 \\
 & $3.60 \pm 0.09$ & 3C 147 \\
$5.0$ & $3.38 \pm 0.05$ & 3C 48 \\
 & $3.50 \pm 0.07$ & 3C 147 \\
 & $3.46 \pm 0.05$ & 3C 286 \\
$5.1$ & $3.16 \pm 0.10$ & 3C 48 \\
 & $3.52 \pm 0.11$ & 3C 147 \\
$5.2$ & $3.34 \pm 0.05$ & 3C 286 \\
$5.5$ & $3.34 \pm 0.05$ & 3C 147 \\
$5.6$ & $3.20 \pm 0.06$ & 3C 147 \\
$6.0$ & $2.96 \pm 0.05$ & 3C 286 \\
$6.1$ & $2.99 \pm 0.10$ & 3C 147 \\
$6.2$ & $2.91 \pm 0.10$ & 3C 147 \\
$6.3$ & $2.98 \pm 0.10$ & 3C 147 \\
$6.4$ & $2.90 \pm 0.10$ & 3C 147 \\
$6.6$ & $2.76 \pm 0.06$ & 3C 147 \\
$6.7$ & $2.86 \pm 0.06$ & 3C 147 

\enddata
\end{deluxetable}

\begin{deluxetable}{ccccc}
\tablecaption{Flux density measurements of NGC~253\label{t:ngc253}}
\tablehead{
  \colhead{$\nu$} & \multicolumn{3}{c}{Flux (Jy)} & \colhead{Calibrator} \\
  (GHz) & \colhead{Core} & \colhead{Extended} & \colhead{Total} & 
}
\startdata
$1.1$ & $2.98 \pm 0.05$ & $3.92 \pm 0.07$ & $6.90 \pm 0.08$ & 3C 147 \\
$1.2$ & $2.93 \pm 0.05$ & $3.83 \pm 0.09$ & $6.75 \pm 0.10$ & 3C 147 \\
$1.3$ & $2.78 \pm 0.04$ & $3.51 \pm 0.06$ & $6.29 \pm 0.07$ & 3C 147 \\
$1.4$ & $2.76 \pm 0.09$ & $3.17 \pm 0.10$ & $5.93 \pm 0.13$ & 3C 48 \\
 & $2.75 \pm 0.06$ & $3.22 \pm 0.07$ & $5.97 \pm 0.09$ & 3C 147 \\
$1.5$ & $2.54 \pm 0.08$ & $2.59 \pm 0.09$ & $5.13 \pm 0.13$ & 3C 48 \\
$1.7$ & $2.48 \pm 0.05$ & $2.68 \pm 0.07$ & $5.16 \pm 0.09$ & 3C 147 \\
$1.8$ & $2.39 \pm 0.05$ & $2.50 \pm 0.07$ & $4.89 \pm 0.08$ & 3C 147 \\
$1.9$ & $2.36 \pm 0.05$ & $2.42 \pm 0.07$ & $4.78 \pm 0.08$ & 3C 147 \\
$2.0$ & $2.35 \pm 0.08$ & $2.29 \pm 0.10$ & $4.64 \pm 0.12$ & 3C 48 \\
$2.1$ & $2.25 \pm 0.08$ & $1.96 \pm 0.09$ & $4.21 \pm 0.12$ & 3C 48 \\
$2.6$ & $1.95 \pm 0.03$ & $1.80 \pm 0.05$ & $3.75 \pm 0.06$ & 3C 147 \\
$2.7$ & $2.01 \pm 0.02$ & $1.74 \pm 0.03$ & $3.75 \pm 0.04$ & 3C 147 \\
$3.0$ & $1.98 \pm 0.07$ & $1.76 \pm 0.09$ & $3.74 \pm 0.12$ & 3C 48 \\
$3.1$ & $1.89 \pm 0.06$ & $1.70 \pm 0.08$ & $3.59 \pm 0.11$ & 3C 48 \\
 & $1.83 \pm 0.02$ & $1.51 \pm 0.03$ & $3.34 \pm 0.04$ & 3C 147 \\
$3.3$ & $1.76 \pm 0.02$ & $1.40 \pm 0.04$ & $3.16 \pm 0.04$ & 3C 147 \\
$3.6$ & $1.75 \pm 0.04$ & $1.38 \pm 0.07$ & $3.13 \pm 0.08$ & 3C 147 \\
$3.7$ & $1.75 \pm 0.08$ & $1.15 \pm 0.17$ & $2.90 \pm 0.19$ & 3C 147 \\
$4.5$ & $1.50 \pm 0.02$ & $0.97 \pm 0.04$ & $2.47 \pm 0.04$ & 3C 147 \\
$4.6$ & $1.46 \pm 0.04$ & $1.06 \pm 0.07$ & $2.51 \pm 0.08$ & 3C 147 \\
$4.7$ & $1.47 \pm 0.03$ & $1.00 \pm 0.09$ & $2.47 \pm 0.10$ & 3C 147 \\
$4.9$ & $1.35 \pm 0.04$ & $1.01 \pm 0.07$ & $2.35 \pm 0.08$ & 3C 147 \\
$5.0$ & $1.37 \pm 0.10$ & $0.65 \pm 0.15$ & $2.01 \pm 0.18$ & 3C 48 \\
 & $1.45 \pm 0.04$ & $0.81 \pm 0.09$ & $2.26 \pm 0.10$ & 3C 147 \\
$5.1$ & $1.34 \pm 0.07$ & $0.90 \pm 0.15$ & $2.23 \pm 0.17$ & 3C 48 \\
$5.6$ & $1.30 \pm 0.03$ & $0.90 \pm 0.09$ & $2.20 \pm 0.10$ & 3C 147 \\
$5.7$ & $1.26 \pm 0.03$ & $0.75 \pm 0.09$ & $2.01 \pm 0.09$ & 3C 147 \\
$5.8$ & $1.26 \pm 0.04$ & $0.61 \pm 0.07$ & $1.87 \pm 0.08$ & 3C 147 \\
$5.9$ & $1.21 \pm 0.04$ & $0.76 \pm 0.08$ & $1.97 \pm 0.09$ & 3C 147 \\
$6.0$ & $1.20 \pm 0.03$ & $0.75 \pm 0.07$ & $1.95 \pm 0.08$ & 3C 147 \\
$6.1$ & $1.16 \pm 0.03$ & $0.60 \pm 0.07$ & $1.76 \pm 0.07$ & 3C 147 \\
$6.3$ & $1.16 \pm 0.03$ & $0.52 \pm 0.07$ & $1.68 \pm 0.08$ & 3C 147 \\
$6.4$ & $1.18 \pm 0.04$ & $0.54 \pm 0.09$ & $1.72 \pm 0.09$ & 3C 147 \\
$6.7$ & $1.13 \pm 0.04$ & $0.73 \pm 0.15$ & $1.87 \pm 0.15$ & 3C 147 \\
$7.0$ & $1.04 \pm 0.04$ & $0.51 \pm 0.07$ & $1.55 \pm 0.09$ & 3C 147 

\enddata
\end{deluxetable}

\begin{deluxetable}{ccc}
\tablecaption{Flux density measurements of Arp~220\label{t:apg220}}
\tablehead{
  \colhead{$\nu$ (GHz)} & \colhead{Flux (Jy)} & \colhead{Calibrator}
}
\startdata
$1.1$ & $0.36 \pm 0.02$ & 3C 48 \\
 & $0.34 \pm 0.01$ & 3C 286 \\
$1.2$ & $0.37 \pm 0.02$ & 3C 48 \\
 & $0.33 \pm 0.01$ & 3C 286 \\
$1.4$ & $0.32 \pm 0.01$ & 3C 286 \\
$1.5$ & $0.26 \pm 0.02$ & 3C 286 \\
$1.6$ & $0.31 \pm 0.03$ & 3C 286 \\
$1.7$ & $0.30 \pm 0.01$ & 3C 286 \\
$1.8$ & $0.31 \pm 0.01$ & 3C 286 \\
$1.9$ & $0.30 \pm 0.01$ & 3C 286 \\
$2.0$ & $0.30 \pm 0.01$ & 3C 286 \\
$2.5$ & $0.27 \pm 0.02$ & 3C 48 \\
 & $0.29 \pm 0.01$ & 3C 286 \\
$2.6$ & $0.28 \pm 0.02$ & 3C 48 \\
 & $0.28 \pm 0.01$ & 3C 286 \\
$2.9$ & $0.27 \pm 0.01$ & 3C 286 \\
$3.2$ & $0.27 \pm 0.01$ & 3C 286 \\
$3.3$ & $0.27 \pm 0.01$ & 3C 286 \\
$3.4$ & $0.28 \pm 0.01$ & 3C 286 \\
$3.5$ & $0.27 \pm 0.01$ & 3C 286 \\
$3.6$ & $0.24 \pm 0.01$ & 3C 286 \\
$4.3$ & $0.25 \pm 0.02$ & 3C 286 \\
$4.4$ & $0.23 \pm 0.02$ & 3C 286 \\
$4.5$ & $0.24 \pm 0.02$ & 3C 286 \\
$4.6$ & $0.24 \pm 0.02$ & 3C 286 \\
$4.7$ & $0.26 \pm 0.02$ & 3C 286 \\
$5.0$ & $0.24 \pm 0.02$ & 3C 48 \\
 & $0.23 \pm 0.01$ & 3C 286 \\
$5.2$ & $0.23 \pm 0.01$ & 3C 286 \\
$5.4$ & $0.24 \pm 0.01$ & 3C 286 \\
$5.7$ & $0.20 \pm 0.02$ & 3C 48 \\
 & $0.22 \pm 0.01$ & 3C 286 \\
$6.0$ & $0.22 \pm 0.01$ & 3C 286 \\
$6.2$ & $0.29 \pm 0.03$ & 3C 286 \\
$6.3$ & $0.20 \pm 0.03$ & 3C 48 \\
 & $0.20 \pm 0.01$ & 3C 286 \\
$6.7$ & $0.23 \pm 0.03$ & 3C 286 \\
$7.0$ & $0.19 \pm 0.02$ & 3C 286 

\enddata
\end{deluxetable}


\begin{deluxetable}{lllccl}
  \tablecaption{High-frequency observations. See \S\ref{s:isotherm}
    for discussion.\label{t:hfobs}}
\tablehead{
  \colhead{$\nu$} &
  \colhead{Flux} &
  \colhead{Telescope} &
  \colhead{Tel. Kind\tablenotemark{a}} &
  \colhead{$\theta_\mathrm{LAS}$\tablenotemark{b}} &
  \colhead{Reference} \\
  (GHz) & (Jy) & & & (arcsec) & 
}
\startdata
\cutinhead{M82}
22.4 & $1.00 \pm 0.15$ & VLA D-cfg. & I & 45 & \citet{ck91} \\
23 & $1.36 \pm 0.07$ & Effelsberg 100m & SD & & \citet{kekw79} \\
24.5 & $1.19 \pm 0.03$ & Effelsberg 100m & SD & & \citet{kwm88} \\
32 & $1.02 \pm 0.06$ & Effelsberg 100m & SD & & \citet{kwm88} \\
42.2 & $0.82 \pm 0.16$ & VLA C-cfg. & I & 43 & \citet{rvzga04} \\
87.2 & $0.51 \pm 0.08$\tablenotemark{c} & Kitt Peak 11m & SD & & \citet{jhm78} \\
92 & $0.59 \pm 0.09$ & BIMA & I & 50 & \citet{ck91} \\
92 & $0.67 \pm 0.10$ & OVRO mm Arr. & I & 55 & \citet{scbb96} \\
\cutinhead{NGC~253 Core Component}
22 & $0.49 \pm 0.05$ & ATCA EW367-cfg. & I & 61 & \citet{rpgsz06} \\
23.7 & $0.56 \pm 0.06$ & VLA DnC-cfg. & I & 60 & \citet{thwnk05} \\
24 & $0.52 \pm 0.05$ & ATCA EW367-cfg. & I & 56 & \citet{owhw05} \\
94 & $0.32 \pm 0.03$ & BIMA A/B/C-cfg. & I & 103 & \citet{pzwls96} \\
\cutinhead{Arp~220}
22.5 & $0.090 \pm 0.006$ & VLA A-cfg. & I & 2 & \citet{zagv96} \\
23.1 & $0.034 \pm 0.007$\tablenotemark{d} & VLA A-cfg. & I & 2 &
  \citet{n88} \\
42.2 & $0.044 \pm 0.004$ & VLA C-cfg. & I & 43 & \citet{r+05} \\
87 & $0.030 \pm 0.006$\tablenotemark{d} & RAINBOW AB-cfg. & I & 8
  & \citet{intok07} \\
87 & $0.035 \pm 0.007$\tablenotemark{d} & IRAM PdBI & I & 40 &
  \citet{r+91} \\
97.2 & $0.061 \pm 0.010$ & IRAM 30m & SD & & \citet{avmgz00}
\enddata
\tablenotetext{a}{The letter ``I'' denotes an interferometer, ``SD'' denotes
  a single dish.}
\tablenotetext{b}{{}Largest angular scale to which the observation is
  sensitive.}
\tablenotetext{c}{Converted to flux density scale of \citet{baars77}.}
\tablenotetext{d}{Uncertainty of 20\% adopted.}
\end{deluxetable}

\begin{deluxetable}{cccc}
  \tablecaption{Results of spectral modeling.\label{t:fitdata}}
  \tablehead{
    \colhead{Parameter\tablenotemark{a}} &
    \colhead{M82} &
    \colhead{NGC~253} &
    \colhead{Arp~220}
  }
  \startdata
  \srcs & 1.21 & 1.02 & 1.17 \\[1em]
$A$ & $\phantom{-}0.93 \pm 0.01$ & $\phantom{-}0.50 \pm 0.01$ & $-0.48^{+0.007}_{-0.011}$ \\
$B$ & $-0.56 \pm 0.02$ & $-0.51 \pm 0.03$ & $-0.01^{+0.03\phn}_{-0.04\phn}$ \\
$C$ & $-0.12 \pm 0.03$ & $-0.16 \pm 0.03$ & $-0.31^{+0.02\phn}_{-0.04\phn}$ \\
$\tau_1$ & $\phantom{-}0.04^{+0.01}_{-0.01}$ & $\phantom{-}0.27 \pm 0.06$ & $ <0.06$ \\[1em]
$\alpha_5$ & $-0.73 \pm 0.04$ & $-0.74 \pm 0.04$ & $-0.44^{+0.01}_{-0.05}$ \\
$\Delta$ & $-0.24 \pm 0.07$ & $-0.33 \pm 0.06$ & $-0.62^{+0.04}_{-0.08}$ \\
$f_\mathrm{th}$ & $\phantom{-}0.06 \pm 0.02$ & $\phantom{-}0.07^{+0.03}_{-0.01}$ & $ <0.02$

  \enddata
  \tablenotetext{a}{Uncertainties in the parameters are marginalized
    1$\sigma$ values determined via Monte Carlo simulations with 10000
    realizations for each source.}
\end{deluxetable}


\onecolumn

\end{document}